\documentclass[a4paper, 10pt]{amsart}

\usepackage[square,sort&compress]{natbib}%numbers,for just numbered

\usepackage[american]{babel}
\usepackage{amsmath,amssymb}
\usepackage{geometry}                % See geometry.jpg to learn the layout options. There are lots.
\usepackage{graphicx}
\usepackage{amssymb}
\usepackage{pdfsync}
\usepackage{graphicx}
\usepackage{amssymb}
\usepackage{enumerate}
\usepackage{xspace}
\usepackage{boxedminipage}
\usepackage{algorithmicx}
\usepackage[ruled]{algorithm}
\usepackage{algpseudocode}
\usepackage{hyperref}

\usepackage{paralist}

\usepackage{varioref}

\usepackage{amssymb,amsfonts,amsmath}

\usepackage[nearskip=-3pt,captionskip=4pt,
            listofformat=subsimple,
            labelformat=simple]{subfig}

\usepackage{graphicx}

\usepackage{pict2e}

\usepackage{commath}
\usepackage[american]{babel}
\usepackage{geometry}                % See geometry.jpg to learn the layout options. There are lots.
\usepackage[parfill]{parskip}    % Activate to begin paragraphs with an empty line rather than an indent
\usepackage{pdfsync}
\usepackage{enumerate}
\usepackage{xspace}
\usepackage{boxedminipage}
\usepackage{hyperref} 

\usepackage{amsthm}

\newcommand{\Proofname}{Proof}
\newenvironment{Proof}[1][{}]
{\noindent{\bf \Proofname\ #1}}
{{\raggedright{{ }\hfill\qed}}}

  \newtheoremstyle{note}% name
  {\parskip}% Space before, vuoto = `valore di default'
  { }% Space after
  {}% body font
  {}% Indent (empty = no indent, \parindent = para indent)
  {\bfseries}% Thm head font
  {}% Punctuation after the heading
  {.5ex}% Space after heading: \newline = to start at next line
  {}% Thm head spec (can be left empty, meaning `normal')

\makeatletter
\renewcommand\subsection{\@startsection
  {subsection}{2}{0mm}%name, level, indent
  {-\baselineskip}%before-skip
  {0pt}%after-skip
  {\bfseries\normalsize}%font specs
}
\makeatother
\newcommand{\mysubsection}[1]{\subsection{#1}}

\usepackage{color}

\swapnumbers
\theoremstyle{note}
\numberwithin{equation}{section}
\newtheorem{The}[subsection]{Theorem} 
\newtheorem{Lem}[subsection]{Lemma}

\newtheorem{Rem}[subsection]{Remark}
\newtheorem{Defn}[subsection]{Definition}
\newtheorem{Assum}[subsection]{Assumption}
\newtheorem{Prop}[subsection]{Proposition}
\newcommand{\Program}[1]{\textsf{#1}\xspace}
\newcommand{\dt}{\ensuremath{\partial_t}}

\newcommand{\myvec}[1]{\ensuremath{\boldsymbol{#1}}}
\newcommand{ \bl}{\color{blue}}

\definecolor{MyGreen}{rgb} {0.05,0.4,0.05}
\definecolor{RedViolet}{rgb} {0.1,0.1,0.75}
\newcommand \gr{\color{MyGreen}}

\usepackage{ifthen}
\provideboolean{showchanges}
\setboolean{showchanges}{false}%%false
\newcommand{\changes}[1]{
  \ifthenelse{\boolean{showchanges}} {{\bl{#1}}} {#1}
}
\provideboolean{refereechanges}
\setboolean{refereechanges}{true}%%true
\newcommand{\refchanges}[1]{
  \ifthenelse{\boolean{refereechanges}} {{\normalsize {#1}}} {#1}
}
\provideboolean{majorrefereechanges}
\setboolean{majorrefereechanges}{true}%%false
\newcommand{\majorrefchanges}[1]{
  \ifthenelse{\boolean{majorrefereechanges}} {{\normalsize {#1}}} {#1}
}
\provideboolean{fullchanges}
\setboolean{fullchanges}{false}%%false
\newcommand{\fullexistence}[1]{
  \ifthenelse{\boolean{fullchanges}} {{\gr{#1}}} {}
}
\provideboolean{shownotes}
\setboolean{shownotes}{false}%%false
\newcommand{\standout}[1]{\colorbox{red}{\textcolor{white}{#1}}}
\newcounter{margnote}[page]
\newcommand{\margnotemark}{{\standout{\footnotesize\upshape\texttt{\arabic{margnote}}}}}
\newcommand{\margnote}[2][]{
  \ifthenelse{
    \boolean{shownotes}
  }{\stepcounter{margnote}\margnotemark\marginpar{
      \texttt{
        \begin{minipage}{2cm}
          \raggedright\tiny
          \margnotemark{#1}: 
          #2
        \end{minipage}
  }}}{}
}
\provideboolean{thesisnotes}
\setboolean{thesisnotes}{false}%%false
\newcommand{\tstandout}[1]{\colorbox{yellow}{\textcolor{white}{#1}}}
\newcounter{tmargnote}[page]
\newcommand{\tmargnotemark}{{\tstandout{\footnotesize\upshape\texttt{\arabic{margnote}}}}}
\newcommand{\thesisnote}[2][]{
  \ifthenelse{
    \boolean{thesisnotes}
  }{\stepcounter{tmargnote}\tmargnotemark\marginpar{
      \texttt{
        \begin{minipage}{2cm}
          \raggedright\tiny
          \tmargnotemark{#1}: 
          #2
        \end{minipage}
  }}}{}
}
\usepackage{upgreek}
\newcommand{\lap}{\ensuremath{\Updelta}}
\renewcommand{\dif}{\operatorname{d}}

\begin{document}
\bibliographystyle{spmpscinat}
  \title{Global existence for semilinear reaction-diffusion systems on evolving domains}
   \date{\today}
  \author{Chandrasekhar Venkataraman \and Omar Lakkis \and Anotida Madzvamuse}
  \address{Department of mathematics\\
	    University of Sussex\\
	    Falmer\\
	    Near Brighton\\
	    UK, BN1 9RF} 
	    \email{c.venkataraman@sussex.ac.uk}
             \urladdr{www.sussex.ac.uk/maths/profile203407}
	    \email{o.lakkis@sussex.ac.uk}
 	    \urladdr{http://www.maths.sussex.ac.uk/Staff/OL/index.html}
	    \email{a.madzvamuse@sussex.ac.uk}
 	    \urladdr{http://www.maths.sussex.ac.uk/~anotida/}
\begin{abstract}
  We present global existence results for solutions of
  reaction-diffusion systems on evolving domains. Global existence
  results for a class of reaction-diffusion systems on fixed domains
  are extended to the same systems posed on spatially linear
  isotropically evolving domains.  The results hold without any
  assumptions on the sign of the growth rate. The analysis is valid
  for many systems that commonly arise in the theory of pattern
  formation. We present numerical results illustrating our theoretical
  findings.
\end{abstract}

 \maketitle

%%%%%%%%%%%%%%%%%%%%%%%%%%%%%%%%%%%%%%%%%%%%%%%%%%%%%%%%%%%%%%%%%%%%%%%%
%%%%%%%%%%%%%%%%%%%%%%%%%%%%%%%%%%%%%%%%%%%%%%%%%%%%%%%%%%%%%%%%%%%%%%%%
%\keywords{Reaction-diffusion systems, global existence, evolving domains, biological pattern formation.} 
%%%%%%%%%%%%%%%%%%%%%%%%%%%%%%%%%%%%%%%%%%%%%%%%%%%%%%%%%%%%%%%%%%%%%%%%

\section{Introduction}
Since their seminal introduction by \citet{turing1952cbm},
reaction-diffusion systems (RDS's) have constituted a standard
framework for the mathematical modelling of pattern formation in
chemistry and biology.  Recent advances in mathematical modelling and
developmental biology identify the important role of \emph{domain
  evolution} as central in the formation of patterns, both empirically
\citep{kondo1995rdw} and computationally
\citep{crampin1999,madzvamuse2007vin,comanici2008pgs}.  In this
respect, many numerical studies, such as
\citet{ano2006} and \citet{barrass2006mtm}, of RDS's on evolving domains are
available. Yet, fundamental mathematical questions such as existence
and regularity of solutions of RDS's on evolving domains remains an
important open question \citep{kelkel2009weak}.

\changes{ 
Numerous studies on the stability of solutions of RDS's on fixed
domains are available, for example,
\citet{rothe1984gsr,hollis1987global,wei2008sms}, but very little
literature regarding the stability of solutions of RDS's on evolving
domains. \citet{madzvamuse2009stability} provides a linear stability
analysis of RDS's on continuously evolving domains, and
\citet{labadie2007stabilizing} examines the stability of solutions of
RDS's on monotonically growing surfaces.  Our discussion here differs
from all these studies in that we focus on planar evolving domains and
 we show existence, uniqueness and
stability, for an entire class of RDS's on evolving
domains independently of the rate of evolution.  In this article we
prove the stability of solutions of RDS's on a particular, but
fundamentally important, class of time-evolving domains: that of
\emph{bounded spatially linear isotropically evolving domains}.
}

\changes{
We show that if a RDS fulfils a restricted version of certain
\emph{stability conditions}, introduced by \citet{morgan1989ges} for
fixed domains, then the RDS fulfils the same stability conditions on
any bounded spatially linear isotropic evolution of the domain.  We
thus prove that, under certain conditions, the existence and
uniqueness for a RDS on a fixed domain implies the existence and
uniqueness for the corresponding RDS on an evolving domain. This is, to
our best knowledge, the first result that holds independently of the
growth rate and is thus valid on growing or contracting domains as
well as domains that exhibit periods of growth and periods of
contraction.  Our analysis rigourously justifies computations
for this type of domain evolution, and we illustrate our results
with benchmark computations using a moving finite element (Lagrangian)
approach.
}

\changes{
The outline of our discussion goes as follows: In \S \ref{S2} we state
our model problem together with the form of domain evolution that we
consider, and present a transformation of our model system to the
Lagrangian framework that helps in proving global existence of
solutions.  In \S \ref{S3} we review the existence results for RDS's on
fixed domains which will form the basis of our analysis.  In
\S \ref{S4} we state and prove the central results of this work; in
particular, we extend the existence results cited in \S \ref{S3} to
problems posed on evolving domains.  In \S \ref{S5} we illustrate some
specific applications of our results, in particular those of
significance in the field of biological pattern formation.} We focus on
growth functions commonly encountered in the field of developmental
biology for which our analysis is valid and show the applicability of
our analysis to some of the important reaction kinetics encountered in
the theory of biological pattern formation. In \S \ref{S6} we present
numerical results for a RDS posed on a periodically evolving
domain. We present a moving finite element scheme and a fixed domain
finite element scheme to approximate the solution of a RDS posed on
the evolving and the Lagrangian frame respectively. In \S \ref{S7} we
summarise our findings and indicate future research directions.
%%%%%%%%%%%%%%%%%%%%%%%%%%%%%%%%%%%%%%%%%%%%%%%%%%%%%%%%%%%%%%%%%%%%%%%% 
\section{RDS's on continuously evolving domains}\label{S2}
Let $\myvec{u}\left(\myvec{x},t\right)$ be a ($m\times1$) vector of
concentrations of chemical species, with $\myvec{x} \in
\mathbb{R}^n$, the time-dependent spatial variable and $t \in [0,T],
\ T>0,$ the time variable. The model problem we wish to consider is a
semilinear RDS posed on a continuously evolving domain  (see
\citet{ano2000} for details of the derivation), given by,
\begin{equation}
  \begin{split}
    \label{eqn:gen}
    \partial_{t}\myvec{u}\left(\myvec{x},t\right)+[\nabla\cdot(\myvec{a}:\myvec{u})]\left(\myvec{x},t\right)&=\myvec{f}(\myvec{u}\left(\myvec{x},t\right))+\myvec{D}\lap\myvec{u}\left(\myvec{x},t\right)  \qquad \mbox{for } \myvec{x} \in \Omega_{t} \mbox{ and } t \in (0,T], 
  \end{split}
\end{equation}
where $\Omega_{t} \subset \mathbb{R}^n\, (\ n<\infty)$ is a $C^{2+\gamma}(\Omega),$ simply connected, bounded and continuously deforming domain with respect to
$t$. The function $\myvec{f}$ is a ($m\times1$) vector of nonlinear coupling terms that is locally Lipschitz,
 $\myvec{D}$ is a diagonal matrix with strictly positive entries on the diagonal,  $\myvec{a}$ is a flow velocity generated by the evolution of the domain and $(\myvec{a}:\myvec{u}) := (\myvec{a}u_1,\dotsc,\myvec{a}u_m)^T$ by definition.
\begin{Assum}[Flow velocity]
\label{a_assum}
 We assume that the flow velocity ${a}_i(\myvec{x},t)$ is identical
 to the domain velocity, i.e., $${a}_i = {\dt{x_i}}\quad i=1, \dotsc, n,$$
 as is standard in the derivation of RDS's on evolving domains on application of Reynold's Transport Theorem \citep{acheson1990elementary}.
 \end{Assum} 
To simplify the exposition we take boundary conditions to be of homogenous Neumann type. \citet{morgan1989ges} considers a much wider class of boundary conditions and our analysis may be extended to this more general setting. We are primarily interested in patterns that arise as a result of self-organisation prompting the consideration of homogenous Neumann boundary conditions. We take the initial condition for each $u_i$  to be  bounded and nonnegative. Our model problem thus takes the form (\ref{eqn:gen}), equipped with the following boundary and initial conditions:
  \begin{equation}\label{eqn:boundary_initial}
 \begin{split}
\begin{cases}
[\vec{\nu}\cdot\nabla\myvec{u}](\myvec{x},t)=0, \quad & \myvec{x} \in \partial\Omega_{t}, t>0,\\
{\myvec{u}}(\myvec{x},0)={\myvec{u}}_{0}(\myvec{x}),  \quad&\myvec{x} \in \Omega_{0}.
\end{cases}
\end{split}
\end{equation}
\mysubsection{Lagrangian transformation}\label{lagtran}
For our analysis it is more convenient to work with problems defined on a time-independent domain. Therefore, we introduce a transformation that maps our model problem (\ref{eqn:gen}) from a time-dependent domain to a fixed domain (see \citet{baines1994mfe} for a more detailed discussion of this approach). In order to do this, without too many technical complications, we will restrict our attention to special evolutions of the domain, described next.
\begin{Assum}[Isotropic domain evolution]  We assume the domain $\Omega_t$ to evolve by obeying a bounded spatially linear isotropic domain deformation, i.e.,
\begin{equation}
 \begin{split}
\label{gfnc}
\myvec{x}=\rho(t)\myvec{\xi} \qquad \text{ for all }  (\myvec{\xi},t) \in\Omega_0\times[0,T] \mbox{ and all } \myvec{x}\in\Omega_t,
\end{split}
\end{equation}
\label{rhodefn}with \majorrefchanges{$\rho \in C^{2}\left([0,T] ; 0,\infty\right)$} and where $\myvec{\xi}$ represents the spatial coordinates of the initial domain.
This assumption and Assumption \ref{a_assum} imply that
\begin{equation}
\begin{split}
\myvec{a}(\myvec{x},t)=\dot{\rho}(t)\myvec{\xi},
\end{split}
\end{equation}
where $\dot{\rho} := \frac{\dif \rho}{\dif t}$. 
\end{Assum} 
Hence, we obtain the following transformed problem on the initial domain (see  \citet{madzvamuse2007vin} for details), with  
\begin{equation}
\hat{\myvec{u}}(\myvec{\xi},t)=\myvec{u}(\rho(t)\myvec{\xi},t) \qquad \mbox{ for }  t \in [0,T] \mbox{ and } \myvec{\xi} \in \Omega_0,
\end{equation} 
we have
 \begin{equation}
 \begin{split}
\label{eqn:tran}
\begin{cases}
\partial_{t}{\myvec{\hat{u}}}+n\frac{\dot{\rho}}{\rho}{\myvec{\hat{u}}}=\myvec{f}({\myvec{\hat{u}}})+\frac{\myvec{D}}{\rho^{2}}{\lap}{\myvec{\hat{u}}} \quad &\mbox{on }\Omega_{0} \times (0,T],\\
[\vec{\nu}\cdot\nabla\hat{\myvec{u}}](\myvec{\xi},t)=0 \quad &\mbox{on } \partial\Omega_{0}, t>0,\\
{\myvec{\hat{u}}}({\myvec{\xi}},0)={\myvec{\hat{u}}}_{0}({\myvec{\xi}})  \quad&{\myvec{\xi}} \in \Omega_0,\\
0 \leq {\myvec{\hat{u}}}_{0}({\myvec{\xi}}) < \infty,&
\end{cases}
\end{split}
\end{equation}
where $ \Omega_{0}$ is the initial spatial domain, $n$ is the spatial dimension and the Laplacian is now taken  with respect to $\myvec{\xi}$.
The local and global existence results that we utilise from the existing literature require the coefficients on our diffusion term to be independent of time; to this end we introduce the following proposition:
\majorrefchanges{
\begin{Prop}[Time rescaling  \citep{labadie2007stabilizing}]\label{prop:tr}
Let $\myvec{u}$ be a solution of $(\ref{eqn:gen})$, rescaling time via the change of variables
\begin{equation}
 \begin{split}
\label{eqn:rescaling}
s(t):=\int_{0}^t\frac{\dif r}{\rho(r)^2},
\end{split}
\end{equation}
and denoting $S:=s(T)$. We have, $\myvec{u}\left(\rho(t)\myvec{\xi},t\right)=\tilde{\myvec{u}}(\myvec{\xi},s)$, where $\tilde{\myvec{u}}$ satisfies  
\begin{equation}
\begin{split}
\label{eqn:scaled}
\begin{cases}
\partial_{s}{\tilde{\myvec{u}}}+n{\rho}{\dot{\rho}}{\tilde{\myvec{u}}}=\rho^{2}\myvec{f}({\tilde{\myvec{u}}})+\myvec{D}{\lap}{\tilde{\myvec{u}}}  \quad &\mbox{on }\Omega_{0} \times (0,S],\\
[\vec{\nu}\cdot\nabla\tilde{\myvec{u}}](\myvec{\xi},s)=0 \quad&{\myvec{\xi}} \in \partial\Omega, s >0,\\
{\tilde{\myvec{u}}}({\myvec{\xi}},0)={\tilde{\myvec{u}}}_{0}({\myvec{\xi}}) \quad&{\myvec{\xi}} \in \Omega.
\end{cases}
\end{split}
\end{equation}
Furthermore, if $\myvec{f}(\myvec{u})$ is locally Lipschitz  in $\myvec{u}$ then $\tilde{\myvec{f}}(\tilde{\myvec{u}}(\vec{\xi},s),s)=\rho^{2}(s)\myvec{f}({\tilde{\myvec{u}}}(\vec{\xi},s))-n{\rho}(s){\dot{\rho}}(s){\tilde{\myvec{u}}}(\vec{\xi},s),$ is locally Lipschitz in $\tilde{\myvec{u}}$.% $ for all $ t\in[0,T]$.
\end{Prop}
\begin{Proof}  We note that with domain evolution of the form considered in this study, there exist  $C_{1}, C_{2}$ such that $0 < \rho \leq C_{1} < \infty$ and $\|\dot{\rho}\|_{L_{\infty}[0,T]} \leq  C_{2} < \infty$. 
Applying the rescaling (\ref{eqn:rescaling}), we see that for any function $g \in C^1[0,T]$
 \begin{equation}
 \begin{split}
\label{eqn:rescaling2}
\partial_{s}g\left(t\right)= \partial_{s}t(s)\partial_{t}g(t)=\rho^{2}(t)\partial_{t}g(t).
\end{split}
\end{equation}
Defining $\tilde{\myvec{u}}(\myvec{\xi},s):=\myvec{u}(\rho(t){\myvec{\xi}},t)$ and multiplying problem (\ref{eqn:tran}) by $ \rho^2 $, we obtain problem (\ref{eqn:scaled}).
 Clearly since $\myvec{f}(\myvec{u})$ is locally Lipschitz in $\myvec{u}$,  Assumption \ref{rhodefn} implies $\tilde{\myvec{f}}(\tilde{\myvec{u}},s)$ is locally Lipschitz in $\tilde{\myvec{u}}$ and globally Lipschitz in $s$. 
\end{Proof}}
\section{Basic theoretical setting for RDS's on fixed domains}\label{S3}
We now summarise the existing results for RDS's on fixed domains which will form the basis of our analysis on evolving domains. \refchanges{The following result is a straightforward generalisation of  \citet[Prop. 1]{hollis1987global}.
\begin{The}[Local existence]\label{locexis}
Let Assumption \ref{gfnc} hold.  Problem (\ref{eqn:scaled}) admits a unique local solution. Furthermore, defining the unique maximal solution of (\ref{eqn:scaled}) by 
\begin{equation}\label{eqn:T_max_defn}
\tilde{\myvec{u}}:\Omega_0\times[0,T_{\max})\to\mathbb{R}^m,
\end{equation}
there exists a function $\myvec{N}\in{C}([0,T_{\max});\mathbb{R}^m)$ such that,
%There exists $T_{\max}>0$ and $\myvec{N}\in{C}([0,T_{\max});\mathbb{R}^m)$ such that, Problem (\ref{eqn:scaled}) admits a unique maximal classical solution $\tilde{\myvec{u}}(\myvec{\xi},t)$ on the interval $[0,T_{\max})$. With
\begin{equation}
\begin{split}
\label{eqn:continuosly_bounded_locexis}
\|\tilde{u}_{i}(\cdot,s)\|_{L_{\infty}(\Omega_0)}\leq{N}_i(s)  \quad\text{ for } i\in[1,\dotsc,{m}]\text{ and } {s}\in[0,T_{\max}).
\end{split}
\end{equation}
Finally, if $T_{\max} < \infty$,
\begin{equation}
\begin{split}
\label{eqn:local}
\lim_{s \to T_{\max}^{-}}\left(\sum_{i=1}^m\|\tilde{u}_{i}(\cdot,s)\|_{L_{\infty}(\Omega_0)}\right)=\infty.
\end{split}
\end{equation}
\end{The}
\mysubsection{Global existence on fixed domains}
%We now present some of the existence results from Morgan \citet{morgan1989ges} on fixed domains.
If $\rho(t)=1$ for all $t\in[0,T]$, Problem (\ref{eqn:scaled}) becomes, find $\tilde{\myvec{u}}:\Omega_0\times(0,T]\to\mathbb{R}^m$ such that,
\begin{equation}
\begin{split}
\label{eqn:fixed}
\begin{cases}
\partial_{t}\tilde{\myvec{u}}=\myvec{f}(\tilde{\myvec{u}})+\myvec{D}{\lap}\tilde{\myvec{u}},  \quad &\mbox{on }\Omega_0 \times (0,T], \\
[\vec{\nu}\cdot\nabla\tilde{\myvec{u}}](\myvec{\xi},t)=0, \quad&{\myvec{\xi}} \in \partial\Omega_0, t>0,\\
\tilde{\myvec{u}}({\myvec{\xi}},0)=\tilde{\myvec{u}}_{0}({\myvec{\xi}}), \quad&{\myvec{\xi}} \in \Omega_0,
\end{cases}
\end{split}
\end{equation}
where we have used the fact that $s(t)=t$ (cf. (\ref{eqn:rescaling})).
\begin{Defn}[Invariant region]
\thesisnote{Can expand on this whole secn. with existence for RDS on fixed domains as a large secn. of the chapter pics for invariant regions etc.}$\Sigma \subset \mathbb{R}^m$ is called an invariant region for the solution of the reaction-diffusion system (\ref{eqn:fixed}) if  for any solution $\myvec{u}$, 
\begin{equation}
\begin{split}
\myvec{u}({\myvec{\xi}},0) \in \Sigma \implies \myvec{u}({\myvec{\xi}},t) \in \Sigma \quad \text{for all} \;  \  t \in (0,T].
\end{split}
\end{equation} 
\end{Defn} 
\begin{Assum}[Positive solutions]  \label{finv}
We assume hereon that 
\begin{equation}
\label{eqn:finv2}
\begin{split} 
{f}_i(\myvec{{u}})|_{{u}_i=0}\geq0 \  \text{for all} \; \ t \in [0,T],
\end{split}
\end{equation} 
and for $\myvec{f}\not \in C^1(\mathbb{R}^m_+;\mathbb{R}^m)$, the strict inequality
\begin{equation}
\label{eqn:finv}
\begin{split} 
{f}_i(\myvec{{u}})|_{{u}_i=0}>0 \ \text{for all} \; \ t \in [0,T].
\end{split}
\end{equation} 
\end{Assum}
Assumption \ref{finv}   together with the positivity of our initial data, implies $\mathbb{R}_+^m$ which we refer to as the positive quadrant,  is  an invariant region for the solutions of problem (\ref{eqn:fixed}) (see \citet[Th.14.7,14.11~pp.200--203]{smoller1994swa}).
%\citet{smoller1994swa} Theorem 14.7 page 200 and Theorem 14.11 page 203).

\begin{Rem}[General invariant regions]\label{rem:finv} Assumption \ref{finv} may be relaxed. The proof of our  existence results only requires bounded initial data and the existence of an invariant region. Furthermore, consideration of the positive quadrant alone is sufficient for our studies. 
\end{Rem}

\mysubsection{Lyapunov stability conditions}\label{subsec:lyapunov}
We now introduce a Lyapunov function for the dynamical system defined by (\ref{eqn:fixed})
 when the initial condition $\tilde{\myvec{u}}_0$ varies which is used to prove global existence and a restricted version of the conditions it is required to fulfil \citep{morgan1989ges}.
  
Suppose $\myvec{f}$ is as defined in problem (\ref{eqn:fixed}) and that there exists a function $H \in C^2\left(\mathbb{R}_{+} ; \mathbb{R}\right)$ and $h_{i} \in C^2\left(\mathbb{R}_{+}; \mathbb{R}\right)$ for each $ i=1,\dotsc, m,$ such that
\label{Gorig}
\begin{equation}
\begin{split} 
H(\myvec{z})=\sum_{i=1}^{m}h_{i}(z_{i})   \quad\text{for all} \; \ \myvec{z} \in \mathbb{R}_{+}^m \label{G1},
\end{split}
\end{equation}
\begin{equation}
\begin{split} 
h_{i}(z_{i}), h_{i}^{\prime\prime}(z_{i}) \geq 0    \quad\text{for all} \; \  \myvec{z} \in \mathbb{R}_{+}^m \label{G2},
\end{split}
\end{equation}
\begin{equation}
\begin{split} 
H(\myvec{z}) \to \infty \iff \myvec{z} \to \infty \quad\text{for all} \; \  \myvec{z} \in \mathbb{R}_{+}^m.\label{G3}
\end{split}
\end{equation}
Suppose there exists $\myvec{A}=(a_{ij}) \in (\mathbb{R})^{m\times{m}}$ satisfying $a_{ij} \geq 0, a_{ii}>0$ with   $1\leq i,j \leq m$ such that for some $r, k_{1}, k_{2} \in \mathbb{R}_{+}$ independent of $j$, we have
\begin{equation}
\begin{split} 
\sum_{i=1}^{j}a_{ij}h_{i}^{\prime}(z_{i})f_{i}(\myvec{z})\leq k_{1}(H(\myvec{z}))^{r} + k_{2}   \quad\text{for all} \; \  \myvec{z} \in \mathbb{R}_{+}^m,   j \leq m.\label{G4}
\end{split}
\end{equation}
Suppose there exist $q,k_{3},k_{4}   \in \mathbb{R}_{+}$ such that for $1 \leq i \leq m$, we have
\begin{equation}
\begin{split} 
h_{i}^{\prime}(z_{i})f_{i}(\myvec{z}) \leq k_{3}(H(\myvec{z}))^q +k_{4},   \quad\text{for all} \; \  \myvec{z} \in \mathbb{R}_{+}^m. \label{G5}
\end{split}
\end{equation}
Suppose there exist $k_{5}, k_{6} \geq 0$ such that 
\begin{equation}
\begin{split} 
\nabla{H}(\myvec{z})\cdot\myvec{f}(\myvec{z}) \leq k_{5}H(\myvec{z}) +k_{6}   \quad\text{for all} \; \ \myvec{z} \in \mathbb{R}_{+}^m. \label{G6}
\end{split}
\end{equation}}
\begin{The}[A priori estimates  \citep{morgan1989ges}]\label{localap}
Let conditions (\ref{G1}), (\ref{G2}) and (\ref{G6}) hold and let $\tilde{\myvec{u}}$ be a solution of problem  (\ref{eqn:fixed}). The following a priori estimates hold,
\refchanges{
\begin{align}
\Big\|\int_{\tau}^{t}H(\tilde{\myvec{u}}(\cdot,s))\dif s\Big\|_{L_{\infty}(\Omega_0)} \leq g(t) \quad   &\mbox{for }  0 \leq\tau<t<T_{\max}, \label{eqn:one_localap}\\
\int_{0}^{t}\int_{\Omega_0} H(\tilde{\myvec{u}}({\myvec{\xi}},s))^{2}\dif{\myvec{\xi}}\dif s \leq \tilde{g}(t) \quad  &\mbox{for } 0\leq{t}<T_{\max}, \label{eqn:two_localap}
\end{align}
where $g,\tilde{g} \in C[0,\infty)$.
}
\end{The}
\begin{The}[Global existence on fixed domains  \citep{morgan1989ges}]\label{globexis}
If  conditions (\ref{G1})---(\ref{G5}) hold, with $r$ from condition (\ref{G4}) satisfying  $r < (1+a)$,  $a\in\mathbb{R}_+$, $\tilde{\myvec{u}}$ is a solution of problem (\ref{eqn:fixed}) and if there exists $g\in C[0,\infty)$ such that
\begin{equation}\label{eqn:globexis_one}
\begin{split}
\Big\|\int_{\tau}^{t}\left\vert{H}\big(\tilde{\myvec{u}}(\cdot,s)\big)\right\vert^a\dif s\Big\|_{L_{\infty}(\Omega_0)} \leq g(t)  \quad  \text{for }  0 \leq\tau<t<T_{\max},
\end{split}
\end{equation}
then $T_{\max} = \infty$. Alternatively, if conditions (\ref{G1})---(\ref{G5}) hold, with $r$ from condition (\ref{G4}) satisfying  $r <  (1+\frac{2b}{n+2})$,   $b>0$ and where $n$ represents the spatial dimension, $\tilde{\myvec{u}}$ solves a problem of the form (\ref{eqn:fixed}) and if there exists $\tilde{g} \in C[0,\infty)$ such that
\begin{equation}\label{eqn:globexis_two}
\begin{split}
\int_{0}^{t}\int_{\Omega_0}\Big|H(\tilde{\myvec{u}}({\myvec{\xi}},s))\Big|^{b}\dif {\myvec{\xi}}\dif s \leq \tilde{g}(t)  \quad  \text{for }0\leq{t}<T_{\max},
\end{split}
\end{equation}
then $T_{\max}=\infty$. 
\end{The}
Specifically if $r$ from condition (\ref{G4}) satisfies  $r < 2$ or if $\Omega_0 \subset \mathbb{R}$   with $ r < \frac{7}{3}$ and the remaining conditions (\ref{G1})---(\ref{G6}) are satisfied then $T_{\max} = \infty$.

\section{Global existence on evolving domains}\label{S4}
\refchanges{In this section we show that, if the stability conditions  in \S \ref{subsec:lyapunov} are valid for Problem (\ref{eqn:fixed}) then they remain valid under any evolution of the domain fulfilling Assumption \ref{rhodefn}, given a suitable assumption on the structure of ${H}$. We also extend the previous a priori estimates and existence results of \citet{morgan1989ges}  to problems with time dependent $\myvec{f}$.}
\begin{Assum}[Polynomial Lyapunov function]
\label{lyap}We assume the Lyapunov function introduced in \S \ref{Gorig} is of the following form
\begin{equation}
\begin{split}
H(\myvec{z})=\sum_{i=1}^{m}z_{i}^{p_{i}} ,\quad  p_{i} \geq 1 \mbox{ for }  i=1,\dotsc,m.
\end{split}
\end{equation}
\end{Assum}

\begin{Rem}[Polynomial growth restriction]
Assumption \ref{lyap} is somewhat natural. Condition (\ref{G5}) is essentially a polynomial type growth restriction on the zero order terms \citep{morgan1989ges}. Assumption \ref{lyap}  can be viewed as the explicit analogue of the polynomial growth restriction on the zero order terms implicit in (\ref{G5}).
\end{Rem}
\begin{Lem}[Equivalence of Lyapunov functions]\label{Lem:equivalence_lyap} %If there exists a function $H(\myvec{z})$ satisfying conditions (\ref{G1})---(\ref{G6}) as defined in {\S} \ref{Gorig}  with $\myvec{f}$ from the fixed domain problem (\ref{eqn:fixed}), $H(\myvec{z})$ is of the form given in Assumption \ref{lyap} and Assumption \ref{finv} holds then $H(\myvec{z})$ satisfies conditions (\ref{G1})---(\ref{G6})  with $\myvec{f}$ replaced by $\tilde{\myvec{f}}$  from the transformed problem  (\ref{eqn:scaled}), $r$ from condition (\ref{G4}) either unchanged or 1 and other constants that may differ.
\refchanges{Suppose Assumptions \ref{rhodefn}, \ref{finv} and \ref{lyap} hold. Let the Lyapunov stability conditions in \S \ref{subsec:lyapunov} be satisfied by $H$ and $\myvec{f}$. Then the conditions in \S \ref{subsec:lyapunov} with  $r$ (cf. (\ref{G4})) replaced by $\tilde{r}:=\max(1,r)$, are satisfied by $H$ and $\tilde{\myvec{f}}$ (cf. (\ref{eqn:f_tilde_defn})) in place of $\myvec{f}$.}
\end{Lem}
\begin{Proof}
We denote the zero order term in Problem (\ref{eqn:scaled}) by 
\begin{equation}\label{eqn:f_tilde_defn}
\begin{split}
\myvec{\tilde{f}}\left(\tilde{\myvec{u}}(\vec{\xi},s),s\right) := \rho^{2}(s)\myvec{f}\left({\tilde{\myvec{u}}}(\vec{\xi},s)\right)-n{\dot{\rho}}(s){\rho}(s){\tilde{\myvec{u}}}(\vec{\xi},s). 
\end{split}
\end{equation}
The positive quadrant  remains an invariant region for the solutions of our evolving domain problem since
\begin{equation}
\begin{split}
{\tilde{f}_i}\left(\tilde{\myvec{u}}(\vec{\xi},s),s\right)|_{u_{i}=0} &= \rho^{2}(s){f_i}\left(\tilde{\myvec{u}}(\vec{\xi},s)\right)|_{u_{i}=0},
\end{split}
\end{equation}
thus Assumption \ref{finv} implies  $\mathbb{R}_+^m$ is an invariant region for the solutions of problem (\ref{eqn:scaled}).
Let  $k_{i}, i=1,\dotsc,6, q, r$  \mbox{and}  $\myvec{A}$ be as defined in \S  \ref{Gorig}, for which conditions (\ref{G1})---(\ref{G6}) hold for problem (\ref{eqn:fixed}). 
Denote ${C}_{1}:= \|\rho\|_{L_{\infty}[0,T]}$ and $C_{2}:= \|\dot{\rho}\|_{L_{\infty}[0,T]}$; these are well defined real numbers thanks to  to Assumption \ref{rhodefn}.  We now show that conditions (\ref{G1})---(\ref{G6})  hold with the same $H$,  $\myvec{{f}}$ replaced by $\myvec{\tilde{f}}$ and $r$ from (\ref{G4}) replaced by $\tilde{r}$, where $\tilde{r} = \max(1,r)$. 

Clearly conditions (\ref{G1})---(\ref{G3}) are still satisfied as they depend only on $H$ which is unchanged. Condition (\ref{G4}) holds since
\begin{equation}
\begin{split}
\sum_{i=1}^{j}a_{ij}h_{i}^{\prime}\tilde{f}_{i}=\sum_{i=1}^{j}a_{ij}h_{i}^{\prime}(\rho^2f_{i}-n\dot{\rho}\rho{\tilde{u}}_i)\leq (k_{1}({H})^{r} + k_{2})C_{1}^2 +nC_{1}C_{2}\sum_{i=1}^{j}a_{ij}h_{i}^{\prime}\tilde{u}_i,
\end{split}
\end{equation}
by the stability of the fixed domain problem. Assumption \ref{lyap} gives,
 \begin{equation}
\begin{split}\label{G4_scaled}
\sum_{i=1}^{j}a_{ij}h_{i}^{\prime}\tilde{f}_{i}&\leq (k_{1}({H})^{r} + k_{2})C_{1}^2 +nC_{1}C_{2}\sum_{i=1}^{j}a_{ij}p_{i}h_i\\ 
&\leq (k_{1}({H})^{r} + k_{2})C_{1}^2 + k_{7}{H}\leq (k_{1}C_{1}^2 +k_{7})({H})^{\tilde{r}} + k_{8}, 
\end{split}
\end{equation}
where $$\tilde{r} = \max(1,r) \ \mbox{and} \  k_{8} : =
    \begin{cases}
      k_{2}C_{1}^2 + k_{1}C_{1}^2,  & \mbox{ if } r < 1,\\
       k_{2}C_{1}^2,  &  \mbox{ if } r  =1,\\
       k_{2}C_{1}^2 + k_{7}, & \mbox{ if } r>1.
    \end{cases}$$
Condition (\ref{G5}) holds since
\begin{equation}
\begin{split}\label{G5_scaled}
h_i^{\prime}\tilde{f}_{i} &\leq C_{1}^2h_i^{\prime}{f}_{i} + nC_{1}C_{2}p_{i}h_i\\
&\leq k_{3}C_{1}^{2}({H})^q + k_4C_{1}^2 + nC_{1}C_{2}\max_{i}(p_i){H}\\ 
&\leq k_{10}{H}^{\tilde{q}} + k_{11},  
\end{split}
\end{equation}
where $$\tilde{q} = \max(1,q) \ \mbox{and}  \  {k}_{11} : =
    \begin{cases}
      k_4C_{1}^2 + k_{3}C_{1}^{2},  & \mbox{ if } q \leq 1,\\
      k_4C_{1}^2, & \mbox{ if } q= 1,\\
       k_{4}C_{1}^2 + k_{9}, & \mbox{ if } q>1.
    \end{cases}$$
Condition (\ref{G6})  holds since
\begin{equation}
\begin{split}\label{G6_scaled}
\nabla{H}.\myvec{\tilde{f}}\leq \sum_{i=1}^mC_{1}^2h_i^{\prime}{f}_{i} + nC_{1}C_{2}p_{i}h_i\leq (k_{5}{H} +k_{6})C_{1}^2 + k_{12}{H}.
\end{split}
\end{equation}

Thus the positive quadrant remains an invariant region for the solutions of problem (\ref{eqn:scaled}) and  the Lyapunov stability conditions in \S \ref{subsec:lyapunov} are satisfied, completing the proof.
\end{Proof}

 \margnote{IMPROVE}
\begin{Rem}[Applicability of \citet{morgan1989ges} to systems with time dependent zero order terms] 
\majorrefchanges{ 
  Suppose the reaction function
  $\tilde{\myvec{f}}\left(\tilde{\myvec{u}}(\myvec{\xi},t),t\right)$
  is locally Lipschitz with respect to $\tilde{\myvec{u}}$ and $t$,
  and suppose that the Lyapunov function $H$ depends only on
  $\tilde{\myvec{{u}}}$. Then, Theorems \ref{localap} and
  \ref{globexis} remain applicable \cite[(5.5)]{morgan1989ges},
  \cite[Th. 1.1]{morgan1995existence} and
  \cite[Th. 4]{bendahmane-mathematical}. Thus, the Lipschitz result
  of Proposition \ref{prop:tr}, the structural Assumption \ref{lyap}
  and the equivalence of Lyapunov functions proved in Lemma
  \ref{Lem:equivalence_lyap} imply that Theorems \ref{localap} and
  \ref{globexis} are applicable for solutions of (\ref{eqn:scaled}).

For completeness, we include a proof of Theorems \ref{localap} and
\ref{globexis} for solutions of Problem (\ref{eqn:scaled}), in
Appendices \ref{appendix:local} and \ref{appendix:global}
respectively.  To remain concise we prove a sufficient existence
result for the examples presented in \S \ref{S5} and briefly sketch
the full proof of Theorems \ref{localap} and \ref{globexis}.

The results of \citet{morgan1995existence} apply for systems with time
dependent diffusion. This may allow treatment of more general domain
evolution where the rescaling carried out in \S \ref{S2} yields a
system with time dependent diffusion. We leave this generalisation for
future studies.  }
\end{Rem}

\begin{The}[Global existence of solutions on evolving domains]\label{finalthm}
\refchanges{
Let Assumptions  \ref{rhodefn}, \ref{finv} and \ref{lyap} hold and suppose  $H$, $\myvec{f}$ and $r$  satisfy the conditions in \S \ref{subsec:lyapunov} with   $r <  2$ or  if $\Omega \subset \mathbb{R},  r < \frac{7}{3}$ (cf. (\ref{G4})).  Then, Problem (\ref{eqn:gen}) admits a global classical solution.
}
\end{The}

\begin{Proof}
\refchanges{
Application of the results in \S \ref{lagtran} allows us to show existence for the transformed Problem  (\ref{eqn:scaled}) defined on a fixed domain. Theorem \ref{locexis} gives the existence of a unique non-continuable classical solution. From Lemma \ref{lyap} the stability conditions in \S \ref{subsec:lyapunov} hold with $\tilde{\myvec{f}}$ (cf. (\ref{eqn:f_tilde_defn})), $H$  and $r <  2$ or  if $\Omega \subset \mathbb{R},$  $r < \frac{7}{3}$.
Theorem \ref{localap} gives an a priori estimate for $H$.  Theorem \ref{globexis} implies  $T_{\max}=\infty$ (cf. (\ref{eqn:T_max_defn})) completing the proof.
}
\end{Proof}

\section{Applications}\label{S5}
In this section we illustrate some applications of Theorem \ref{finalthm}. We present different forms of admissible domain evolution that fufil Assumption \ref{rhodefn}. We show that  Assumption \ref{lyap} is applicable to some  commonly encountered models in chemistry and biology. We concentrate on RDS's which admit Turing instabilities, as the main focus of our research is biological pattern formation. We identify and describe Lyapunov  functions and constants that imply global existence for the fixed domain problem and thus for the evolving domain problem by Theorem \ref{finalthm}.
\mysubsection{Admissible domain evolution}
We now provide some commonly encountered examples of domain evolution in developmental biology for which Assumption \ref{rhodefn} holds:
\begin{itemize}
\item Logistic evolution on any finite positive time interval
\begin{equation}
\rho(t)=\frac{e^{r_g{t}}}{1+\frac{1}{K}(e^{r_g {t}}-1)},  \quad t \in [0,T],
\end {equation}
where $r_g \geq 0$ is the growth rate and $K>1$ is the carrying capacity (limiting size of the evolving domain).
\item Exponential evolution on any finite positive time interval 
\begin{equation}
\rho(t)={e^{r_g t}}, \quad t \in [0,T].
\end {equation}
\item Linear evolution on any finite positive time interval 
\begin{equation}
\rho(t)={1+{r_g t}}, \quad t \in [0,T],
\end {equation}
where $r_g > -\frac{1}{T}$.
\end{itemize}
%%%%%%%%%%%%%%%%%%%%%%%%%%%%%%%%%%%%%%%%%%%%%%%%%%%%%%%%%%%%%%%%%%%%%%%%
\mysubsection{Admissible kinetics}
We now present some of the commonly encountered reaction kinetics of problem (\ref{eqn:gen}) for which the analysis of \citet{morgan1989ges} implies global existence of solutions on fixed domains. 
We first consider the problem (\ref{eqn:gen}) with this general reaction term 
\begin{equation}\label{eqn:oscillator_kinetics}
\begin{split}
f_{i}(\myvec{u})&= \sum_{j=1}^{m}c_{ij}u_{j} + (-1)^{i}g(\myvec{u}) + b_{i},
\end{split}
\end{equation}
with the following restrictions
\begin{align}
c_{ij}&  \geq 0, \quad \mbox{for}\quad i \neq j. \label{C1}\\
b_{i}& \geq0 ,  \quad  i = 1,\dotsc,m. \label{C2}\\
 g(\myvec{u})|_{u_{i}=0}& = 0,    \quad  i = 1,\dotsc,m. \label{C3}\\
g(\myvec{u}) & \leq \left(\sum_{i=1}^{m}u_{i}\right)^p,  \quad  \text{for all} \; \  \myvec{u} \in \mathbb{R}_{+}^m. \label{C4}\\
g& \in C^1(\mathbb{R}_+^m;\mathbb{R}).  \label{C5}
\end{align}
\changes{
%%This sentence, though important, is not worth so much pomp as a ``Remark''.
%%\begin{Rem}[Physical motivation]
  The motivation of this type of kinetics, as discussed by
  \citet{murray2003mathematical}, is their role in the theory of
  biological oscillators due to a feedback mechanism.
%%\end{Rem}
}
\changes{
\begin{Prop}[Lyapunov function]\label{prop:lyapunov_function}
We show that problem (\ref{eqn:gen}) equipped with kinetics (\ref{eqn:oscillator_kinetics}) is well
posed, with Lyapunov function ${H(\myvec{z})} :=
\sum_{i=1}^{m}z_{i}$. 
\end{Prop}
}
\begin{Proof}
\changes{Recalling that the initial data is bounded} and nonnegative (\ref{eqn:boundary_initial}), we show
that Assumption \ref{finv} is fulfilled, which implies $\mathbb{R}^+_m$
is an invariant region for the solutions.\changes{Indeed, from} (\ref{C3}) we have
\begin{equation}
  \begin{split}
    f_{i}(\myvec{u})|_{u_{i}=0}=\sum_{j=1}^{m}c_{ij}u_{j} +
    b_{i}
    =
    \sum_{j=1}^{i-1}c_{ij}u_{j} + \sum_{j=i+1}^{m}c_{ij}u_{j} +
    b_{i}.
  \end{split}
\end{equation}
Conditions (\ref{C1}) and (\ref{C2}) imply
\begin{equation}
  \begin{split}
    f_{i}(\myvec{u})|_{u_{i}=0}
    \geq
    b_{i}\geq0 \quad \text{for all}
    \quad \ \myvec{u} \in \mathbb{R}_{+}^m.
  \end{split}
\end{equation}
Thus Assumption \ref{finv} is fulfilled due to (\ref{C5}).  Now we
show conditions (\ref{G1})---(\ref{G6}) are fulfilled with $r = 1$.
Clearly conditions (\ref{G1})---(\ref{G3}) hold.
Condition (\ref{G4}) holds
with
\begin{equation*}
\begin{split}
a_{ij} :=\begin{cases}
      1 \quad \mbox{if} \ j = i, \\
      1 \quad  \mbox{if} \ j = 1 \ \mbox{and} \ i \ \mbox{is even,}\ \\
      0  \quad \mbox{otherwise,}
         \end{cases}
         \end{split}
         \end{equation*} 
$k_1 = \max_{i,j}(c_{ij})$, $k_2 = \sum_{i=1}^{m}b_i $ and $r = 1$.
Condition (\ref{G5}) holds since 
\begin{equation}
\begin{split}
h_i^{\prime}{f}_{i}&= f_i\leq \max_{i,j}(c_{ij}) \sum_{i=1}^{m}u_i + \max_{i}(b_{i}) + g(\myvec{u}).
\end{split}
\end{equation}
Using (\ref{C4}) we have
\begin{equation}
\begin{split}
h_i^{\prime}{f}_{i}&\leq \max_{i,j}(c_{ij}) \sum_{i=1}^{m}u_i + \max_{i}(b_{i}) + (\sum_{i=1}^{m}u_{i})^p \\
&\leq \max_{i,j}(c_{ij}){H(u)} + {H(u)}^p + \max_{i}(b_{i})\leq k{H(u)}^q + b,
\end{split}
\end{equation}
where $k, q, b \in \mathbb{R}_+$ represent constants that depend on the value of $p$ in (\ref{C4}). 
Condition (\ref{G6}) holds since
\begin{equation}
\begin{split}
\nabla{H}\cdot\myvec{{f}} &= \sum_{i=1}^{m}f_i\leq \max_{i,j}(c_{ij}) \sum_{i=1}^{m}u_i + \sum_{i=1}^{m}b_{i}\leq k{H(u)}+b,\\
\end{split}
\end{equation}
where $k = \max_{i,j}(c_{ij})$ and $b = \sum_{i=1}^{m}b_{i}$.
A straightforward application of Theorem \ref{finalthm} thus completes the proof.  
\end{Proof}
%%%%%%%%%%%%%%%%%%%%%%%%%%%%%%%%%%%%%%%%%%%%%%%%%%%%%%%%%%%%%%%%%%%%%%%%
\subsection{Examples}
The generic problem for which we showed global existence of solutions actually encompasses some of the more widely studied models in the theory of pattern formation such as the Gray-Scott model and the Brussellator. Below we present two examples of two species reaction terms for which our analysis implies global existence of solutions, the first of which is a restriction of the reaction term above.
\begin{itemize}
\item{{\it Activator-depleted} substrate model:}
We consider the {\it activator-depleted} substrate model \citep{schnakenberg1979simple,gierer72,prigo68}  also known as
the Brusselator model:
\begin{equation}\label{eqn:schnak}
\begin{split}
\begin{cases}
f_1\left(u_{1}, u_{2}\right)&=\gamma\left(a-u_{2}^2u_{1}\right),\\
f_2\left(u_{1}, u_{2}\right)&=\gamma\left(b-u_{2}+u_{2}^2u_{1}\right),
\end{cases}
\end{split}
\end{equation}
where $0 < a, b, \gamma < \infty $.
The assumption of nonnegative initial data implies that the positive quadrant is invariant for our problem due to the fact that $a, b > 0$.  
If we take $H(\myvec{u})=u_1 + u_2,$ then conditions (\ref{G1})---(\ref{G6}) are fulfilled with $r=0$ which implies global existence of solutions on evolving domains via Theorem \ref{finalthm}. The remaining constants for which conditions (\ref{G1})---(\ref{G6}) hold are given in Table \ref{tab:const}.

\item{Thomas reaction kinetics:}
The following model, proposed and studied experimentally by \citet{thomas1975artificial},  is based on a specific reaction involving the substrates oxygen and uric acid which react in the presence of the enzyme uricase: 
\begin{equation}\label{eqn:thom}
\begin{split}
\begin{cases}
f_1\left(u_{1}, u_{2}\right)&=\gamma\left(a-u_{1}- g(u_{1},u_{2})\right),\\
f_2\left(u_{1}, u_{2}\right)&=\gamma\left(b - \alpha{u}_{2}-g(u_{1},u_{2})\right),
\end{cases}
\end{split}
\end{equation}
where
\begin{equation}
\begin{split}
g(u_{1},u_{2}) = \frac{\kappa{u}_{1}u_{2}}{1+u_{1}+\beta{u}_{1}^2},
\end{split}
\end{equation}
and $0 < \gamma, a, \alpha, b, \kappa,  \beta < \infty$. Once again the assumption of nonnegative initial data  implies that the positive quadrant is invariant for our problem due to the fact that $a, b > 0$. If we again take $H(\myvec{u})=u_1+u_2,$  then conditions (\ref{G1})---(\ref{G6}) are fulfilled with $r=0$ which implies global existence of solutions on evolving domains via Theorem \ref{finalthm}. The remaining constants for which conditions (\ref{G1})---(\ref{G6}) hold are given in Table \ref{tab:const}.
\end{itemize}

\begin{Rem}[Invariant rectangles for the Thomas model] It can be shown, utilising the techniques of \citet{smoller1994swa}, that there exist bounded invariant rectangles for the solutions of the Thomas model defined in (\ref{eqn:thom}). This implies global existence of solutions via Theorem \ref{locexis}. However the authors can show that it is possible to construct  growth functions that fulfil Assumption \ref{rhodefn}, for which the bounded invariant rectangle can be made arbitrarily large.  This necessitates the Lyapunov function approach to show existence and uniqueness of solutions.
\end{Rem}

\begin{table}[htdp]
\begin{center}
\begin{tabular}{|c|c|c|}
 \hline
 Parameters&{\it Activator-depleted} substrate model&Thomas model\\
 \hline
 $a_{11}$&$1$&$1$\\ 
 $a_{12}$&$0$&$0$\\
 $a_{21}$&$1$&$0$\\
  $a_{22}$&$1$&$1$\\
  $k_1$&$0$&$0$\\
  $k_2$&$\gamma(a+b)$&$ \gamma(a+b)$\\
  $k_3$&$\frac{\gamma}{2}$&$0$\\
  $q$&$3$&$0$\\
  $k_4$&$\gamma(\max(a,b))$&$\gamma(\max(a,b))$\\
  $k_5$&$0$&$0$\\
  $k_6$&$\gamma(a+b)$&$\gamma(a+b)$\\
    \hline
    \end{tabular}
\end{center}
\caption{Terms from \S \ref{S3} for which conditions (\ref{G1})---(\ref{G6}) hold for the kinetics defined in (\ref{eqn:schnak}) and (\ref{eqn:thom}) respectively.}
\label{tab:const}
\end{table}%

\begin{Rem}[Further applications] The analysis can be applied to a large number of problems unrelated to the theory of pattern formation. \citet{garvie-analysis} provide an example applicable to ecology  and the aforementioned paper of \citet{morgan1989ges} contains further examples as well as the numerous citations of said paper that use the approach on various problems.
\end{Rem}

\section{Numerical experiments}\label{S6}
In this section we present numerical results on two-dimensional evolving domains to back-up the theoretical results of the previous sections. The numerical simulation of RDS's on growing domains is an extensive research area. \citet{crampin2002mda}  and \citet{madzvamuse2007vin} study mode doubling and tripling behaviour of the {\it activator-depleted} substrate model  on one- and two-dimensional growing domains.  The theoretical results derived above apply for any evolution that fulfils Assumption \ref{rhodefn}. We present numerical results on a periodically evolving domain that exhibit spot splitting as well as spot annihilation and merging, which to the authors knowledge is as yet an unstudied area.

\mysubsection{Domain evolution}
We consider periodic domain evolution  defined by 
\begin{equation}
 \rho(t)=1+9\sin\left(\frac{\pi{t}}{T}\right),  \quad t \in [0, 1000=T], 
\end{equation}
 with the initial domain defined as  $\Omega_{0}=[-0.25,0.25]^2$  which grows to $\Omega_{500}=[-2.5,2.5]^2$  before contracting back to the original domain. 

\mysubsection{Continuous problems}
We present results for the aforementioned  {\it activator-depleted} model considered on a periodically evolving domain and its equivalent transformed system on a fixed domain as in \S \ref{lagtran}. For example, on a periodically evolving domain the problem is stated as follows: 
 \begin{equation}
\label{eqn:schnak_moving}
\begin{cases}
\partial_{t}{u_1}(\myvec{x},t)+[\nabla\cdot(\myvec{a}{u_1})](\myvec{x},t)-{\lap}{u_1}(\myvec{x},t)=0.1-[u_{2}^2u_{1}](\myvec{x},t), \quad &\mbox{for } \myvec{x}\in\Omega_{t}\\
\partial_{t}{u_2}(\myvec{x},t)+[\nabla\cdot(\myvec{a}{u_2})](\myvec{x},t)-0.01{\lap}{u_2}(\myvec{x},t)=0.9-u_{2}(\myvec{x},t)+[u_{2}^2u_{1}](\myvec{x},t), &\mbox{and } t\in(0,T], \\
[\vec{\nu}\cdot\nabla\myvec{u}](\myvec{x},t)=0, \quad  &\myvec{x} \in \partial\Omega_{t},  t>0,
\end{cases}
\end{equation}
where $\myvec{x}=\rho(t)\myvec{\xi}$ and $a_i(\myvec{x},t)=\dot{\rho}(t)\myvec{\xi},$  $i=1,2.$   Equivalently, the following transformed equations are obtained on a fixed domain,
\begin{equation}
\label{eqn:schnak_fixed}
\begin{cases}
\partial_{t}{u_1}+2\frac{\dot{\rho}}{\rho}{u_1}-\frac{1}{\rho^2}{\lap}{u_1}=0.1-u_{2}^2u_{1}, \quad &\\
\partial_{t}{u_2}+2\frac{\dot{\rho}}{\rho}{u_2}-\frac{0.01}{\rho^2}{\lap}{u_2}=0.9-u_{2}+u_{2}^2u_{1},\quad &\mbox{on }\Omega_{0}\times (0,T], \\
[\vec{\nu}\cdot\nabla\myvec{u}](\myvec{\xi},t)=0, \quad  &{\myvec{\xi}} \in \partial\Omega_{0}, t>0.
\end{cases}
\end{equation}
In both cases we take identical initial conditions as small perturbations around the homogenous steady state of (1.0, 0.9) obtained in the absence of domain evolution. 

\mysubsection{Numerical schemes}\label{subsec:numerical_schemes}
We employ a Galerkin finite element method for the spatial approximation and  an implicit-explicit modified backward Euler scheme for the time integration. Discretising in time we divide the time interval $[0, T ]$ into a partition of N uniform subintervals, $0 = t_0 < \dotsc < t_N = T$ and denote by $\tau := t_{n} - t_{n-1}$ the time step. For the spatial discretisation we  introduce a regular triangulation $\mathcal{T}^0$ of $\Omega_{0}$ with $K\in \mathcal{T}^0$ an open simplex.  We define the following shorthand for a function of time,  $f (t_n ) =: f^n$. 

We define the finite element space on the initial domain $\mathbb{V}^0 \subset H^1(\Omega_{0})$ as,
\begin{equation}
\begin{split}
\label{eqn:fe_space_fixed}
\mathbb{V}^0:=\lbrace\Phi \in H^1(\Omega_{0}):\Phi|_K\in\mathbb{P}^1 \ \text{for all} \; \ K \in \mathcal{T}^0\rbrace,
\end{split}
\end{equation}  
where $\mathbb{P}^1$  denotes  the space of polynomials no higher than degree 1.
For the numerical simulation of equation (\ref{eqn:schnak_moving}) we require finite element spaces defined on the evolving domain. We construct the finite element spaces $\mathbb{V}^n$  according to the following relation between the basis functions  of $\mathbb{V}^n$ and $\mathbb{V}^0$.
\begin{equation}
\begin{split}
\label{eqn:basis}
\Psi^n=\Psi(\rho^n{\myvec{\xi}},t_n)=\Phi({\myvec{\xi}}) \quad n=1,\dotsc,N. 
\end{split}
\end{equation}  
Thus the family of finite element spaces on the evolving domain $\mathbb{V}^n \subset H^1(\Omega_{t_n}) \quad n=1,\dotsc,N$ may be defined as,
\begin{equation}
\begin{split}
\label{eqn:fe_space_moving}
\mathbb{V}^n:=\lbrace\Psi^n \in H^1(\Omega_{t_n}):\Psi^n|_K\in\mathbb{P}^1 \ \text{for all} \; \ K \in \mathcal{T}^n\rbrace,
\end{split}
\end{equation}  
where we have used the fact that the domain evolution is linear with respect to space. We approximate the initial conditions in both schemes by 
\begin{equation}
\begin{split}
\label{eqn:discrete_initial}
\myvec{U}^0=\mathcal{I}\myvec{u}_0({\myvec{\xi}}) \quad \text{for all} \; \ {\myvec{\xi}} \in \Omega_{0},
\end{split}
\end{equation}
where $\mathcal{I}$ is the standard Lagrange interpolant.
 The finite element scheme to approximate the solution to equation (\ref{eqn:schnak_moving}) aims to find $U_1^n, U_2^n  \in \mathbb{V}^n, n= 1, \dotsc, N$ such that
\begin{equation}
\begin{split}
\begin{cases}
\label{eqn:discrete_schnak_moving}
\frac{1}{\tau}{\langle{U_1^n},\Psi^n\rangle} +\langle\nabla{U_1^n},\nabla\Psi^n\rangle= \langle0.1-\left(U^{n-1}_{2}\right)^2U_1^n,\Psi^n\rangle+\frac{1}{\tau}\langle{U_1^{n-1}},\Psi^{n-1}\rangle ,\\
\frac{1}{\tau}{\langle{U_2^n},\Psi^n\rangle} +0.01\langle\nabla{U_2^n},\nabla\Psi^n\rangle=\langle0.9-U_2^n+U^{n-1}_{2}U^{n}_{1}U_2^n,\Psi^n\rangle+\frac{1}{\tau}\langle{U_2^{n-1}},\Psi^{n-1}\rangle,
\end{cases}
\end{split}
\end{equation}
for all $\Psi^n \in \mathbb{V}^n, n= 1, \dotsc, N$. Similarly the finite element scheme to approximate the solution to equation (\ref{eqn:schnak_fixed}) aims to find $W_1^n, W_2^n  \in \mathbb{V}^0, n= 1, \dotsc, N$ such that
\begin{equation}
\begin{split}
\begin{cases}
\label{eqn:discrete_schnak_fixed}
\frac{1}{\tau}{\langle{W_1^n-W_1^{n-1}},\Phi\rangle} +\frac{1}{\left(\rho^n\right)^2}\langle\nabla{W_1^n},\nabla\Phi\rangle + \frac{2\dot{\rho}^n}{\rho^n}\langle{W_1^n},\Phi\rangle= \langle0.1-\left(W^{n-1}_{2}\right)^2W_1^n,\Phi
\rangle,\\
\frac{1}{\tau}{\langle{W_2^n-W_2^{n-1}},\Phi\rangle} +\frac{0.01}{\left(\rho^n\right)^2}\langle\nabla{W_2^n},\nabla\Phi\rangle + \frac{2\dot{\rho}^n}{\rho^n}\langle{W_2^n},\Phi\rangle=\langle0.9-W_2^n+W^{n-1}_{2}W^{n}_{1}W_2^n,\Phi\rangle,
\end{cases}
\end{split}
\end{equation}
for all $\Phi \in \mathbb{V}^0$.  

We solved the models in C utilising the FEM library \Program{ALBERTA}
by \citet{schmidt2005design}. We used the conjugate gradient solver to
compute our discrete solutions. We took an initial triangulation
$\mathcal{T}^0$ with 8321 nodes, a uniform mesh diameter of ${2^{-6}}$
and a fixed timestep of $10^{-3}$. PARAVIEW was used to display our
results.

\begin{figure}[ht]
\includegraphics[trim = 50mm 160mm 25mm 10mm,  clip,  scale=1.4]{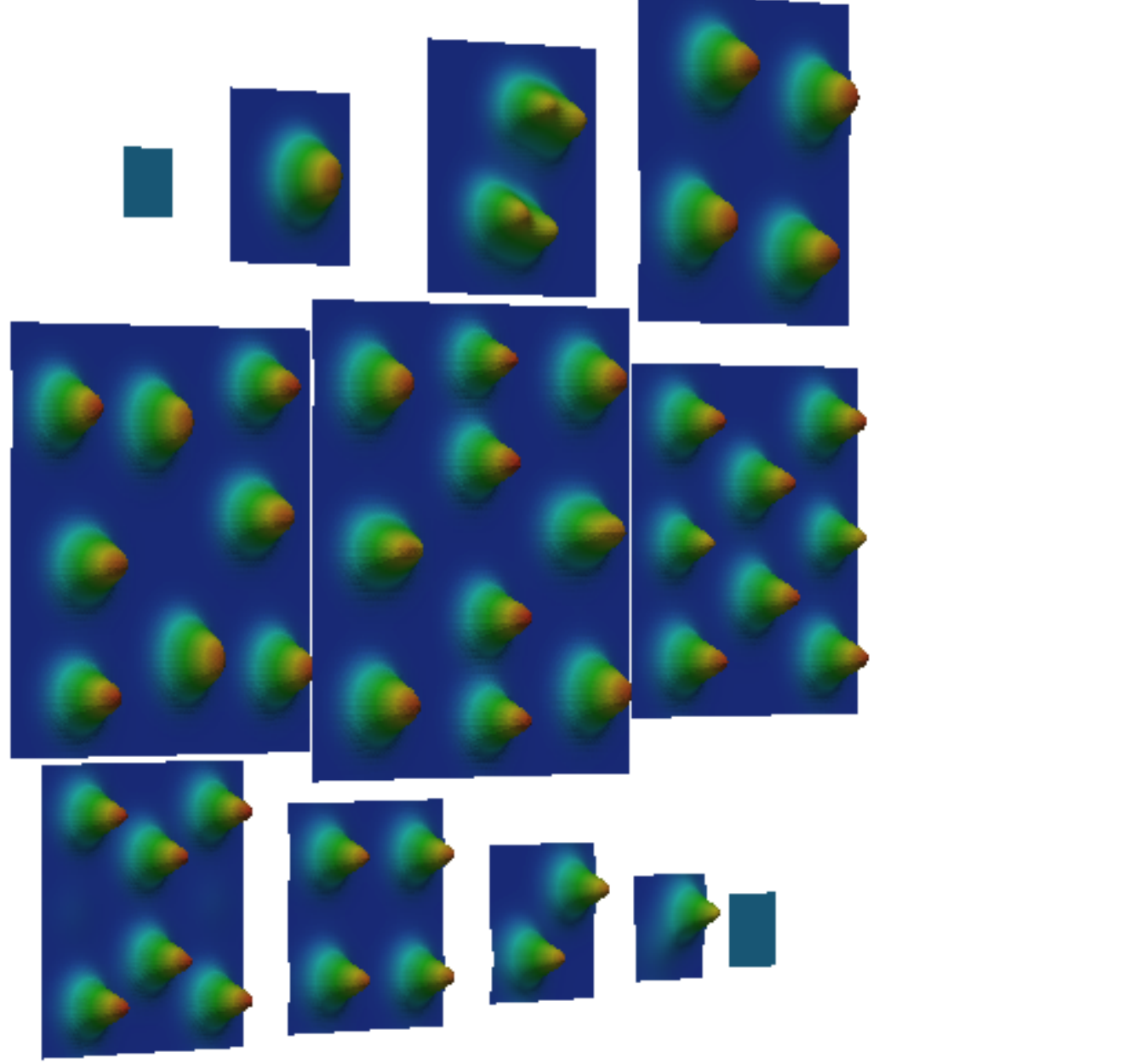}
\caption{Snapshots of the discrete solution $U_2$ corresponding to system (\ref{eqn:discrete_schnak_moving}) at times  0, 50, 160, 220, 380, 500, 700, 740, 820, 900, 980 and 1000 reading from left to right and then top to bottom. For parameter and numerical values see \S \ref{subsec:numerical_schemes}. The solution exhibits a mode doubling sequence of 1,2,4, 8 and finally 10 as the domain grows. As the domain contracts the spots are annihilated in a sequence of 8, 6, 4, 2 and the final transition to a single spot occurs via merging, with the final domain exhibiting no patterns.} 
\label{fig:discrete_schnak_moving}
\end{figure}

\begin{figure}[ht]
\includegraphics[trim = 15mm 40mm 0mm 15mm,  clip,  scale=0.75]{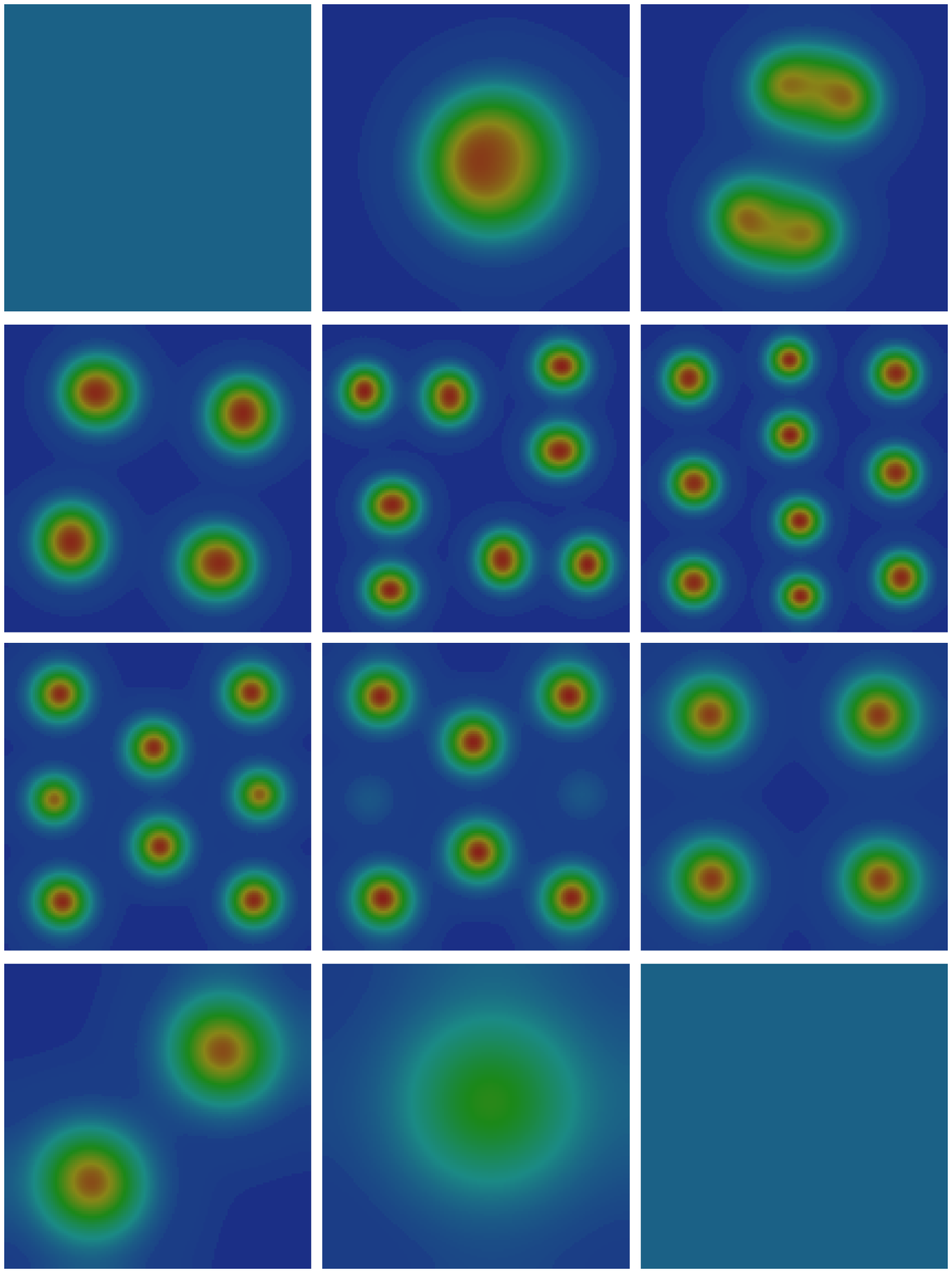}
\caption{Snapshots of the discrete solution $W_2$ corresponding to system (\ref{eqn:discrete_schnak_fixed}) at times 0, 50, 160, 220, 380, 500, 700, 740, 820, 900, 980 and 1000 reading from left to right and then top to bottom. For parameter and numerical values see \S \ref{subsec:numerical_schemes}. The mode transition follows exactly that of Figure \ref{fig:discrete_schnak_moving}, corroborating the results in \S \ref{lagtran}.} 
\label{fig:discrete_schnak_fixed}
\end{figure}

\mysubsection{Results}
Figures \ref{fig:discrete_schnak_moving} and \ref{fig:discrete_schnak_fixed} show snapshots of the activator profile corresponding to the {\it activator-depleted} system (\ref{eqn:schnak}). The inhibitor profiles have been omitted as they are $180^{\circ}$ out of phase to the activator profiles. We have verified numerically that there is very little difference between the discrete solution corresponding to system (\ref{eqn:discrete_schnak_moving})  mapped to the fixed domain and  the discrete solution corresponding to system  (\ref{eqn:discrete_schnak_fixed}) defined on a fixed domain,   as is expected from the results in \S \ref{lagtran}.  

\begin{figure}[ht]
\centering
\includegraphics[trim = 15mm 140mm 40mm 15mm,  clip,  scale=0.8]{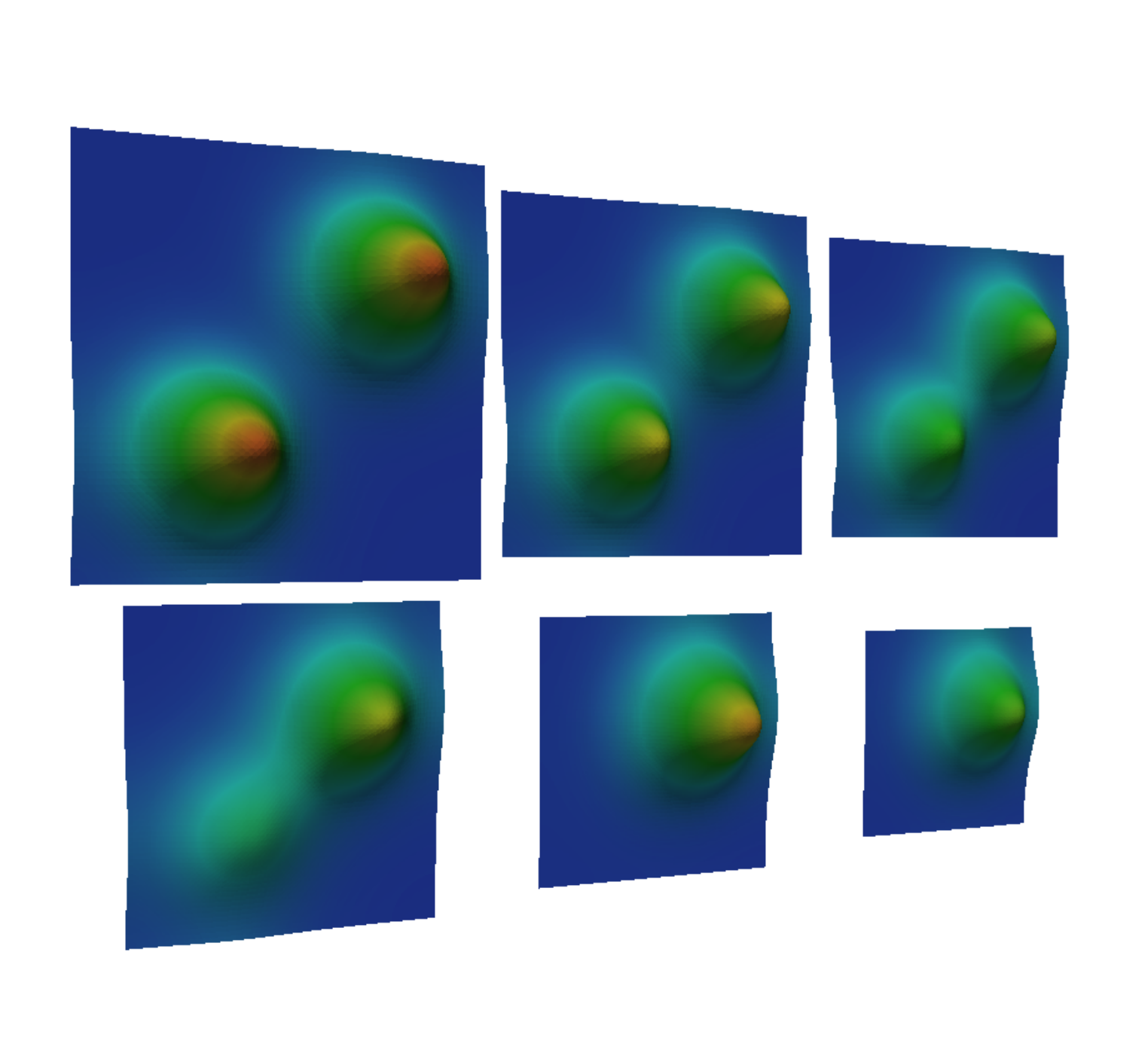}
\includegraphics[trim = 0mm 140mm 0mm 20mm,  clip,  scale=0.55]{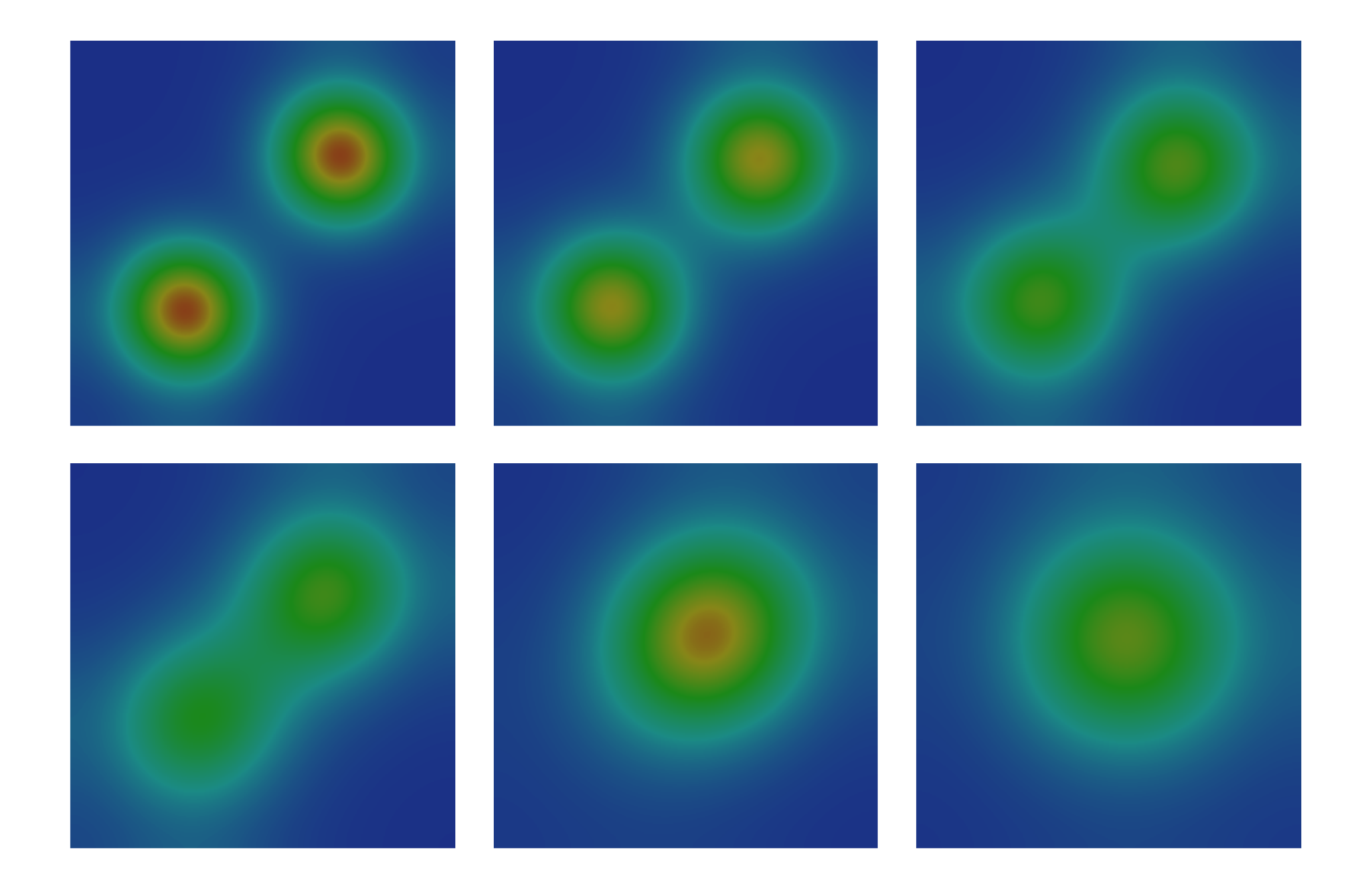}
\caption{ Snapshots of the discrete solution corresponding to system (\ref{eqn:discrete_schnak_moving}) (top) and system (\ref{eqn:discrete_schnak_fixed}) (bottom) at times 930, 940, 950, 955, 965, 975. The spot merging phenomena observed in the transition from two spots to one spot is displayed.} 
\label{fig:discrete_schnak_merging}
\end{figure}

The figures illustrate the mode doubling phenomena that occurs as the domain grows as well as the spot annihilation and spot merging phenomena that occurs as the domain contracts.\majorrefchanges{In Figure \ref{fig:discrete_schnak_merging} we present in more detail the novel spot merging phenomena observed on the contracting domain. It is still unclear whether the spot merging phenomenon is in fact a special case of the spot annihilation phenomenon, that occurs when the modes are of sufficient proximity to influence each other.} 

We note that the mode transition sequence, i.e., the number of spots,
is different when the domain grows to when it contracts. The
difference in the mechanism of mode transitions on growing and
contracting domains is an area in which very little work has been done
and these initial numerical results indicate the need for further
exploration of this area.

\section{Conclusion}\label{S7}
Many problems in biology and biomedicine  involve growth. In developmental biology recent advances in experimental data collection allow experimentalists to capture the emergence of pattern structure formation during growth development of the organism or species. Such experiments include the formation of spot patterns on the surface of the eel, patterns emerging on the surface of the Japanese flounder and butterfly wing patterns forming during the growth development of the imaginal wing disc. In all these examples, patterns form during growth development.  

Since the seminal paper by \citet{turing1952cbm} which considered linear models that could give rise to spatiotemporal solutions on fixed domains due the process of diffusion-driven instability, a lot of theoretical results on global existence of such solutions have been derived and proved for highly nonlinear mathematical models \citet{smoller1994swa,rothe1984gsr}. Only recently, mathematical models on growing domains have been derived from first principles in order to incorporate the effects of domain evolution into the models \citet{ano2000,crampin1999}.  In all these studies, very little analysis has been done up to now to extend the theoretical global existence results to models defined on evolving domains.  

Under suitable assumptions, we have extended existence results from problems posed on fixed domains to problems posed on an evolving domain.  We  have illustrated the applicability of the existence results of \citet{morgan1989ges} to problems on evolving domains. We  have shown that global existence of solutions to many commonly encountered RDS's  on fixed domains implies global existence of solutions to the same RDS's on a class of evolving domains. The results are significant in the theory of pattern formation especially in fields such as developmental biology where problems posed on evolving domains are commonly encountered.  Our results hold with no assumptions on the sign of the growth rate, which may prove useful in other fields where monotonic domain growth is not valid from a modelling perspective.  The applicability of our results is demonstrated by considering different forms of domain evolution (linear, logistic and exponential). 

In order to validate our theoretical findings, we presented results on a periodically evolving domain. Our results illustrate the well-known
period-doubling phenomenon during domain growth but more interesting and surprising is the development of spot annihilation and spot merging phenomena during contraction. This raises new questions about bifurcation analysis on growing and contracting domains. 

One of our primary goals is the numerical analysis of finite element approximations of RDS's on evolving domains. The classical existence results obtained will be an important tool in future work. Numerical experiments have been carried out and they illustrate the need for further numerical analysis especially in the case of contracting domains. Extension of our work onto domains with more complex evolution  is another area for future research.

\section*{Acknowledgments}
The research of C.Venkataraman is partially supported by an EPSRC doctoral training grant and a University of Sussex graduate teaching assistantship. 
\appendix
\clearpage
\section{A priori estimates for systems with time dependent $f$}\label{appendix:local}
\majorrefchanges{
In this section we prove Theorem \ref{localap} remains applicable for solutions of Problem (\ref{eqn:scaled}). For conciseness we focus on assertion (\ref{eqn:one_localap}). We adapt the proof of \citet[Th. 3.2]{morgan1989ges} to our purposes. In Appendix \ref{appendix:global}, we use the a priori estimates obtained in this section to prove the global existence of classical solutions to Problem (\ref{eqn:scaled}). We state the main result of this section in the following Lemma.
 
 \begin{Lem}[An a priori estimate for solutions of Problem (\ref{eqn:scaled})]\label{Lem:apriori_f_tilde}
 Suppose Assumptions \ref{rhodefn}, \ref{finv} and \ref{lyap} hold and let (\ref{G1})---(\ref{G3}) and (\ref{G6}) hold. Then,  assertion (\ref{eqn:one_localap}) is valid with $\tilde{\myvec{u}}$ a solution of Problem (\ref{eqn:scaled}).
 \end{Lem}
 \begin{Proof}
 Let $\tilde{\myvec{f}}$ be as defined in (\ref{eqn:f_tilde_defn}). From the proof of Lemma \ref{lyap},   (\ref{G6_scaled}) holds.
 We split the remainder of the proof into steps.
\begin{enumerate}[Step 1:]
%%%%%%%%%%%%%%%%%%%%%%%%%%%%%%%%%%%%%%%%%%%%%%%%%%%%%%%%%%%%%%%%%%%%%%%%%%%%%%%%%%%%%%%%%%%%%%%%%%%%%%%%%%%%%%
\item
We first show the following inequality for $H$:
 \begin{equation}\label{eqn:localap_int_time}
 \begin{split}
 {H}\left(\tilde{\myvec{u}}(\vec{\xi},t)\right)\leq&\int_{\tau}^t\sum_{i=1}^mD_i \lap_{\vec{\xi}} h_i\left(\tilde{{u}}_i(\vec{\xi},r)\right)+\left(k_5C_1^2+k_{12}\right)H\left(\tilde{\myvec{u}}(\vec{\xi},r)\right)\dif r\\
 &+{H}\left(\tilde{\myvec{u}}(\vec{\xi},\tau)\right)+k_{6}C_1^2(t-\tau) \text{ for } \myvec{\xi}\in\Omega_0\text{ and }t<T_{\max}. 
 \end{split}
 \end{equation}
 From (\ref{eqn:scaled}) we have for $(\myvec{\xi},t)\in\Omega_0\times(0,T_{\max})$,
  \begin{equation}\label{eqn:nabla_H_expression}
 \nabla{H}\left(\tilde{\myvec{u}}(\vec{\xi},t)\right)\cdot\partial_t\tilde{\myvec{u}}(\vec{\xi},t)=\nabla{H}\left(\tilde{\myvec{u}}(\vec{\xi},t)\right)\cdot\left(\myvec{D}\lap\tilde{\myvec{u}}(\myvec{\xi},t)+\tilde{\myvec{f}}\left(\tilde{\myvec{u}}(\myvec{\xi},t),t\right)\right).
 \end{equation}
Using (\ref{G1}),  (\ref{G6_scaled}) and (\ref{eqn:nabla_H_expression}) we obtain the following generalisation of  \citet [(3.1)]{morgan1989ges},
 \begin{equation}\label{eqn:localap_G6}
 \nabla{H}\left(\tilde{\myvec{u}}(\vec{\xi},t)\right)\cdot\partial_t\tilde{\myvec{u}}(\vec{\xi},t)\leq\sum_{i=1}^mD_i\lap_{\vec{\xi}}h_i\left(\tilde{{u}}_i(\vec{\xi},t)\right)+\left(k_5C_1^2+k_{12}\right)H\left(\tilde{\myvec{u}}(\vec{\xi},t)\right)+k_{6}C_1^2,
  \end{equation}
 where we have used the convexity of $H$ (\ref{G2}).
Integrating (\ref{eqn:localap_G6}) in time gives (\ref{eqn:localap_int_time}).
%%%%%%%%%%%%%%%%%%%%%%%%%%%%%%%%%%%%%%%%%%%%%%%%%%%%%%%%%%%%%%%%%%%%%%%%%%%%%%%%%%%%%%%%%%%%%%%%%%%%%%%%%%%%%%
\item
 We use (\ref{eqn:localap_int_time}) to construct an appropriate barrier function with a view to applying the maximum principle, corresponding to \citet[(3.2)---(3.4)]{morgan1989ges}.  Introducing an arbitrary  $T^*<T_{\max}$,  we define 
\begin{align}\label{w_def}
  w(\myvec{\xi},t)&:=\int_\tau^t\sum_{i=1}^m\frac{D_i}{{D^*}}h_i\left(\tilde{u}_i(\myvec{\xi},r)\right)\dif r \text{ for }\vec{\xi}\in\Omega_0 \text{ and } 0\leq\tau<t\leq T^*,
\end{align}
where  $ {D^*}:={\displaystyle\max_i({D_i})}$.
Observe that from (\ref{G1}) and (\ref{w_def}), we have for $0\leq\tau<t\leq T^*$
 \begin{equation}\label{eqn:w_H_inequality}
  \begin{split}
\Bigg\|\int_{\tau}^{t}H(\tilde{\myvec{u}}(\cdot,s))\dif s\Bigg\|_{L_{\infty}(\Omega_0)}&=\Bigg\|\int_{\tau}^{t}\sum_{i=1}^mh_i(\tilde{{u}}_i(\cdot,s))\dif s\Bigg\|_{L_{\infty}(\Omega_0)}\\
&\leq\max_i\left(\frac{D^*}{{D_i}}\right)\|w(\cdot,s)\dif s\|_{L_{\infty}(\Omega_0)}. 
\end{split}
 \end{equation}
From (\ref{eqn:localap_int_time}) and (\ref{w_def}) we obtain  
\begin{equation}
\begin{split}
\begin{cases}\label{eqn:maximum_principle_w}
\partial_tw(\myvec{\xi},t)\leq{D^*}\lap w(\myvec{\xi},t)+L+Mw(\myvec{\xi},t)+k_{6}C_1^2(t-\tau),   &\vec{\xi}\in\Omega_0, 0\leq\tau<t\leq T^*\\
[\myvec{\nu}\cdot\nabla{w}](\myvec{\xi},t)=0,  &\myvec{\xi}\in\partial\Omega_0, t\in(\tau,T^*]\\
w(\myvec{\xi},\tau)=0, &\myvec{\xi}\in\Omega_0,
\end{cases}
\end{split}
\end{equation}
where 
 \begin{align}
 L:= \|H\left(\tilde{\myvec{u}}(\cdot,\tau)\right)\|_{L_{\infty}(\Omega_0)}\text{ and }{M}:=\frac{(k_{12}+C_1^2k_{5}){D^*}}{\displaystyle \min_i{\left(D_i\right)}}.
 \end{align}
Note we have used (\ref{eqn:scaled}) to obtain the boundary conditions. For the purposes of applying the maximum principle we  define a barrier function
 \begin{align}\label{what_def}
  \hat{w}(\myvec{\xi},t)&:=w(\myvec{\xi},t)-\frac{L+K_{6}C_1^2T^*}{M} \text{ for }\vec{\xi}\in\Omega_0 \text{ and } 0\leq\tau<t\leq T^*.
\end{align}
From (\ref{eqn:maximum_principle_w}) and (\ref{w_def}) we have
\begin{equation}
\begin{split}
\begin{cases}\label{eqn:maximum_principle_what}
\partial_t\hat{w}(\myvec{\xi},t)\leq{D^*}\lap \hat{w}(\myvec{\xi},t)+M\hat{w}(\myvec{\xi},t)   &\vec{\xi}\in\Omega_0, 0\leq\tau<t\leq T^*\\
[\myvec{\nu}\cdot\nabla\hat{w}](\myvec{\xi},t)=0,  &\myvec{\xi}\in\partial\Omega_0, t\in(\tau,T^*]\\
\hat{w}(\myvec{\xi},\tau)\leq0, &\myvec{\xi}\in\Omega_0.
\end{cases}
\end{split}
\end{equation}
  %%%%%%%%%%%%%%%%%%%%%%%%%%%%%%%%%%%%%%%%%%%%%%%%%%%%%%%%%%%%%%%%%%%%%%%%%%%%%%%%%%%%%%%%%%%%%%%%%%%%%%%%%%%%%% 
\item 
 We use the maximum principle to complete the proof. Applying the strong maximum principle for parabolic problems \citep[Th. 2.9, Rem. (a)~ pg. 21]{sperb1981maximum}  to (\ref{eqn:maximum_principle_what}) and noting the positivity of $w$, we have
 \begin{align}\label{eqn:w_hat_range}
 -\frac{L+K_{6}C_1^2T^*}{M}\leq\hat{w}(\myvec{\xi},t)\leq{0}\text{  for all }\vec{\xi}\in\Omega_0\text{ and for }t\in(\tau,T^*].
  \end{align}
 From (\ref{what_def}) and (\ref{eqn:w_hat_range}) we have
  \begin{align}\label{eqn:w_max_principle_bounds}
0\leq\hat{w}(\myvec{\xi},t)\leq  \frac{L+K_{6}C_1^2T^*}{M}\text{  for all }\vec{\xi}\in\Omega_0\text{ and for }t\in(\tau,T^*].
  \end{align}
 We conclude from (\ref{eqn:w_max_principle_bounds}) that  
 \begin{align}\label{eqn:w_linf_estimate}
 \|w(\cdot,t)\|_{L_{\infty}(\Omega_0)}\leq\frac{L+K_{6}C_1^2T^*}{M} \text{ for }t\in(\tau,T^*].
 \end{align}
  \end{enumerate}
   Since $T^*$ was arbitrary, combining (\ref{eqn:w_H_inequality}) and (\ref{eqn:w_linf_estimate}) completes the proof of the Lemma.
  \end{Proof}
  %%%%%%%%%%%%%%%%%%%%%%%%%%%%%%%%%%%%%%%%%%%%%%%%%%%%%%%%%%%%%%%%%%%%%%%%%%%%%%%%%%%%%%%%%%%%%%%%%%%%%%%%%%%%%% 

For completeness, we  sketch the proof of assertion (\ref{eqn:two_localap}) with $\tilde{\myvec{u}}$ a solution of Problem (\ref{eqn:scaled}). In (\ref{eqn:localap_G6}) we denote $K_7:=k_5C_1^2+k_{12}$ and $K_8:=k_6C_1^2$,  where $K_7,K_8$ correspond to the terms on the right hand side of \citet[(3.1)]{morgan1989ges}. Assertion (\ref{eqn:two_localap})  follows from the proofs of \citet[Th. 3.3 and 3.4]{morgan1989ges}.  
}
%%%%%%%%%%%%%%%%%%%%%%%%%%%%%%%%%%%%%%%%%%%%%%%%%%%%%%%%%%%%%%%%  %%%%%%%%%%%%%%%%%%%%%%%%%%%%%%%%%%%%%%%%%%%%%%%%%%%%%%%%%%%%%%%%%%%%%%%%%%%%%%%%%%%%%%%%%%%%%%%%%%%%%%%%%%%%%% 
%%%%%%%%%%%%%%%%%%%%%%%%%%%%%%%%%%%%%%%%%%%%%%%%%%%%%%%%%%%%%%%%  %%%%%%%%%%%%%%%%%%%%%%%%%%%%%%%%%%%%%%%%%%%%%%%%%%%%%%%%%%%%%%%%%%%%%%%%%%%%%%%%%%%%%%%%%%%%%%%%%%%%%%%%%%%%%% 
%%%%%%%%%%%%%%%%%%%%%%%%%%%%%%%%%%%%%%%%%%%%%%%%%%%%%%%%%%%%%%%%  %%%%%%%%%%%%%%%%%%%%%%%%%%%%%%%%%%%%%%%%%%%%%%%%%%%%%%%%%%%%%%%%%%%%%%%%%%%%%%%%%%%%%%%%%%%%%%%%%%%%%%%%%%%%%% 
%%%%%%%%%%%%%%%%%%%%%%%%%%%%%%%%%%%%%%%%%%%%%%%%%%%%%%%%%%%%%%%%  %%%%%%%%%%%%%%%%%%%%%%%%%%%%%%%%%%%%%%%%%%%%%%%%%%%%%%%%%%%%%%%%%%%%%%%%%%%%%%%%%%%%%%%%%%%%%%%%%%%%%%%%%%%%%% 
%%%%%%%%%%%%%%%%%%%%%%%%%%%%%%%%%%%%%%%%%%%%%%%%%%%%%%%%%%%%%%%%  %%%%%%%%%%%%%%%%%%%%%%%%%%%%%%%%%%%%%%%%%%%%%%%%%%%%%%%%%%%%%%%%%%%%%%%%%%%%%%%%%%%%%%%%%%%%%%%%%%%%%%%%%%%%%% 
%%%%%%%%%%%%%%%%%%%%%%%%%%%%%%%%%%%%%%%%%%%%%%%%%%%%%%%%%%%%%%%%  %%%%%%%%%%%%%%%%%%%%%%%%%%%%%%%%%%%%%%%%%%%%%%%%%%%%%%%%%%%%%%%%%%%%%%%%%%%%%%%%%%%%%%%%%%%%%%%%%%%%%%%%%%%%%% 
%%%%%%%%%%%%%%%%%%%%%%%%%%%%%%%%%%%%%%%%%%%%%%%%%%%%%%%%%%%%%%%%  %%%%%%%%%%%%%%%%%%%%%%%%%%%%%%%%%%%%%%%%%%%%%%%%%%%%%%%%%%%%%%%%%%%%%%%%%%%%%%%%%%%%%%%%%%%%%%%%%%%%%%%%%%%%%% 
%%%%%%%%%%%%%%%%%%%%%%%%%%%%%%%%%%%%%%%%%%%%%%%%%%%%%%%%%%%%%%%%  %%%%%%%%%%%%%%%%%%%%%%%%%%%%%%%%%%%%%%%%%%%%%%%%%%%%%%%%%%%%%%%%%%%%%%%%%%%%%%%%%%%%%%%%%%%%%%%%%%%%%%%%%%%%%% 
%%%%%%%%%%%%%%%%%%%%%%%%%%%%%%%%%%%%%%%%%%%%%%%%%%%%%%%%%%%%%%%%  %%%%%%%%%%%%%%%%%%%%%%%%%%%%%%%%%%%%%%%%%%%%%%%%%%%%%%%%%%%%%%%%%%%%%%%%%%%%%%%%%%%%%%%%%%%%%%%%%%%%%%%%%%%%%% 
   \fullexistence{
   We split the proof of (\ref{eqn:two_localap}) into two parts. We first show that there exists $g_1\in{C}[0,\infty)$ such that
   \begin{equation}\label{eqn:H_L1bound}
   \|H(\tilde{\myvec{u}}(\cdot,t)\|_{L_1(\Omega_0)}\leq{g_1}(t) \text{ for } 0<t<T_{\max}.
   \end{equation}
   Integrating (\ref{eqn:localap_int_time}) over $\Omega_0$ and setting $\tau=0$ we have
   \begin{equation}\label{eqn:L1_bound_gronwall's}
   \begin{split}
   \|H\left(\tilde{\myvec{u}}(\cdot,t)\right)\|_{L_1(\Omega_0)}\leq\left(k_5C_1^2+k_{12}\right)\|H\left(\tilde{\myvec{u}}\right)\|_{L_1\left([0,t]\times\Omega_0\right)}+|\Omega_0|\left(\|{H}(\tilde{\myvec{u}}_0)\|_{L_{\infty}(\Omega_0)}+k_{6}C_1^2t\right),
 \end{split}
 \end{equation}
 where we have used integration by parts and (\ref{eqn:boundary_initial}) to deal with the first term on the left of (\ref{eqn:localap_int_time}). A Gronwall's argument applied to (\ref{eqn:L1_bound_gronwall's})  implies (\ref{eqn:H_L1bound}). For $0\leq\tau<T_{\max}$, we introduce $z_i$, $i=1,\dotsc,m+1,$  the solution of 
\begin{equation}
\begin{split}
\begin{cases}\label{eqn:maximum_principle_z}
\partial_tz_i(\myvec{\xi},t)={D}_i\lap z_i(\myvec{\xi},t)+k_i(\myvec{\xi},t)  &\vec{\xi}\in\Omega_0, t\in(\tau,T_{\max})\\
[\myvec{\nu}\cdot\nabla{z_i}](\myvec{\xi},t)=0,  &\myvec{\xi}\in\partial\Omega_0, t\in(\tau,T_{\max})\\
z_i(\myvec{\xi},\tau)=l_i, &\myvec{\xi}\in\Omega_0,
\end{cases}
\end{split}
\end{equation}
where for $i=1,\dotsc,m$,
\begin{align*}
k_i&=h_i^\prime\left(\tilde{u}_i(\myvec{\xi},t)\right)\tilde{f}_i\left(\tilde{\myvec{u}}(\myvec{\xi},t)\right),\\
 l_i&=\|h_i\left(\tilde{u}_i(\cdot,\tau)\right)\|_{L_{\infty}(\Omega_0)},
 \end{align*} 
 and  where
 \begin{align*}
D_{m+1}&=D_m,\\
 k_{m+1}&=\left(k_5C_1^2+k_{12}\right)H\left(\tilde{\myvec{u}}(\myvec{\xi},t)\right)+k_{6}C_1^2-\nabla{H}\left(\tilde{\myvec{u}}(\vec{\xi},t)\right)\cdot\tilde{\myvec{f}}\left(\tilde{\myvec{u}}(\myvec{\xi},t),t\right),\\
 l_{m+1}&=l_{m}.
 \end{align*}   
 For $i=1,\dotsc,m$ defining $v_i:=h_i-z_i$ we have from (\ref{eqn:scaled}) and (\ref{eqn:maximum_principle_z})
 \begin{equation}
\begin{split}
\begin{cases}\label{eqn:maximum_principle_v}
\partial_tv(\myvec{\xi},t)={D}_i\left(\lap v_i(\myvec{\xi},t)-h_i^{\prime\prime}\left(\tilde{u}_i(\myvec{\xi},t)\right)\left\vert\nabla\tilde{u}_i(\myvec{\xi},t)\right\vert^2\right)  &\vec{\xi}\in\Omega_0, t\in(\tau,T_{\max})\\
[\myvec{\nu}\cdot\nabla{v_i}](\myvec{\xi},t)=0,  &\myvec{\xi}\in\partial\Omega_0, t\in(\tau,T_{\max})\\
v_i(\myvec{\xi},\tau)=h_i\left(\tilde{u}_i(\myvec{\xi},\tau)\right)-l_i\leq{0} &\myvec{\xi}\in\Omega_0.
\end{cases}
\end{split}
\end{equation}
Thus from (\ref{G2}) and the SMP we have 
\begin{align}\label{eqn:zi>ui}
z_i(\myvec{\xi},t)\geq{h}_i\left(\tilde{u}_i(\myvec{\xi},t)\right)\text{ for } i=1,\dotsc,m.
\end{align}
Summing (\ref{eqn:maximum_principle_z}) over $i$, using  (\ref{eqn:zi>ui}) and applying the
 SMP  we conclude that, there exists $g_2\in{C}[0,T_{\max})$ such that for $i=1,\dotsc,m+1,$ 
 \begin{align}
 0\leq\|z_i(\cdot,t)\|_{L_{\infty}(\Omega_0)}\leq{g_2}(t)\text{ for } \tau\leq{t}<T_{\max}.
 \end{align}
   Furthermore from (\ref{eqn:maximum_principle_z})  we note for each $1\leq{i}\leq{m}, z_i$ satisfies the same $L_1(\Omega_0)$ bound as $h_i$ in (\ref{eqn:H_L1bound}).
We now proceed as in (\ref{w_def}), setting ${\displaystyle{D^*}>\max_{i}(D_i)}$ and defining 
\begin{align}\label{w_def}
  \tilde{w}(\myvec{\xi},t)&:=\int_\tau^t\sum_{i=1}^m\frac{D_i}{{D^*}}z_i\left(\tilde{u}_i(\myvec{\xi},r)\right)\dif r \text{ for }\vec{\xi}\in\Omega_0 \text{ and } 0\leq\tau<t<T_{\max},
\end{align}
we have
\begin{equation}
\begin{split}
\begin{cases}\label{eqn:maximum_principle_w_tilde}
\partial_t\tilde{w}(\myvec{\xi},t)={D^*}\lap \tilde{w}(\myvec{\xi},t)+\sum_{i=1}^{m+1}\left(z_i(\myvec{\xi},\tau)+\left(\frac{D_i}{{D^*}}-1\right)z_i(\myvec{\xi},t)\right)&\\
\quad \qquad \qquad+(k_5C_1^2+k_{12})\int_{\tau}^tH\left(\tilde{\myvec{u}}(\myvec{\xi},r)\right)\dif r+k_{6}C_1^2(t-\tau),   &\vec{\xi}\in\Omega_0, 0\leq\tau<t<T_{\max}\\
[\myvec{\nu}\cdot\nabla{\tilde{w}}](\myvec{\xi},t)=0,  &\myvec{\xi}\in\partial\Omega_0, t\in(\tau,T_{\max})\\
\tilde{w}(\myvec{\xi},\tau)=0, &\myvec{\xi}\in\Omega_0.
\end{cases}
\end{split}
\end{equation}
Thus as in (\ref{eqn:maximum_principle_w}), (\ref{eqn:one_localap}) and the SMP imply that there exists $g_3\in{C}[0,\infty)$ such that 
 \begin{align}
 \|\tilde{w}(\cdot,t)\|_{L_{\infty}(\Omega_0)}\leq{g}_3(t) \text{ for } 0<t<T_{\max}.
 \end{align}
From (\ref{eqn:maximum_principle_w_tilde}), we have for $0\leq\tau<t<T_{\max}$ and $1\leq{i}\leq{m+1}$,
\begin{equation}
\begin{split}
\int_{\tau}^t\int_{\Omega_0}z_i(\myvec{\xi},r)\sum_{j=1}^{m+1}\left(\frac{D_j}{{D^*}}-1\right)z_j(\myvec{\xi},r)\dif\myvec{\xi}\dif r=&\int_{\tau}^t\int_{\Omega_0}z_i(\myvec{\xi},r)\Biggr(\Bigg[\partial_tw-{D^*}\lap{w}-\sum_{i=1}^{m+1}z_i\Bigg](\myvec{\xi},r)\\
&-\left(k_5C_1^2+k_{12}\right)\int_{\tau}^rH\left(\tilde{\myvec{u}}(\myvec{\xi},q)\right)\dif q+k_6C_1^2(r-\tau)\Biggr)\dif \myvec{\xi}\dif r\\
=&\int_{\tau}^t\int_{\Omega_0}-[w{D^*}\lap{z_i}](\myvec{\xi},r)+z_i(\myvec{\xi},r)\Biggr(\Bigg[\partial_tw-\sum_{i=1}^{m+1}z_i\Bigg](\myvec{\xi},r)\\
&-\left(k_5C_1^2+k_{12}\right)\int_{\tau}^rH\left(\tilde{\myvec{u}}(\myvec{\xi},q)\right)\dif q+k_6C_1^2(r-\tau)\Biggr)\dif \myvec{\xi}\dif r\\
\end{split}
\end{equation}

 Defining $k_5C_1^2+k_{12}:=K_7, k_6C_1^2:=K_8$, where $K_7,K_8$ are the terms on the right hand side of \citet[(3.1)]{morgan1989ges} and replicating the proofs of \citet[Th. 3.2, 3.3 and 3.4]{morgan1989ges} we conclude 
\begin{align}
\Big\|\int_{\tau}^{t}H(\tilde{\myvec{u}}(\cdot,s))\dif s\Big\|_{L_{\infty}(\Omega_0)} \leq g(t) \quad   &\mbox{for }  0 \leq\tau<t<T_{\max},\\
\int_{0}^{t}\int_{\Omega_0} H(\tilde{\myvec{u}}({\myvec{\xi}},s))^{2}\dif{\myvec{\xi}}\dif s \leq \tilde{g}(t) \quad  &\mbox{for } 0\leq{t}<T_{\max},
\end{align}
where $g,\tilde{g} \in C[0,\infty)$. Since (\ref{G5_scaled}) holds with $r<2$ or if $\Omega_0\subset\mathbb{R}, r<\frac{7}{3}$, condition (\ref{G4_scaled}) and a duality argument \citet[Th. 2.2 and 2.4 ~ 1137--1139, 1141-1142]{morgan1989ges}   imply that there exists $M_p,N_p>0$ and $0<\delta_p<1$ such that
\begin{equation}
\|H(\tilde{\myvec{u}})\|_{L_p(\Omega_0\times[0,T])}\leq{M}_p+N_p\|H(\tilde{\myvec{u}})\|_{L_p(\Omega_0\times[0,T])}^{\delta_p},
\end{equation}
where $1\leq{p}<\infty$. A bootstrapping argument \citet[Lem. 4.2 ]{morgan1989ges} gives $L_\infty$ bounds which completes the proof. 
}
%%%%%%%%%%%%%%%%%%%%%%%%%%%%%%%%%%%%%%%%%%%%%%%%%%%%%%%%%%%%%%%%%%%%%%%%%%%%%%%%%%%%%%%%%%%%%%%%%%%%%%%%%%%%%%%%%%%%%%%%%%%%%%%%%%%%%%%%%%%%%%%%%%%%%%%%%%%%%%%%%%%%%%%%%%%%%%%%%%%%%%%%%%%%%%%%%%%%%%%%%%%%%%%%%%%%%%%%%%%%%%%%%%%%%%%%%%%%%%%%%%%%%%%%%%%%%%%%%%%%%%%%%%%%%%%%%%%%%%%%%%%%%%%%%%%%%%%%%%%%%%%%%%%%%%%%%%%%%%%%%%%%%%%%%%%%%%%%%%%%%%%%%%%%%%%%%%%%%%%%%%%%%%%%%%%%%%%%%%%%%%%%%%%%%%%%%%%%%%%%%%%%%%%%%%%%%%%%%%%%%%%%%%%%%%%%%%%
\section{Global existence results for systems with time dependent $f$ }\label{appendix:global}
\majorrefchanges{The main result of this section is  Theorem \ref{Thm:existence_time_dependent_f}, a special case of Theorem \ref{globexis}. It is applicable to solutions of Problem (\ref{eqn:scaled}). Theorem   \ref{Thm:existence_time_dependent_f} is enough for our purposes as seen in the examples in \S \ref{S5}. To prove the Theorem, we will modify the proof of \citet[Th. 2.2]{morgan1989ges} with stronger control of the parameter $r$, that appears in (\ref{G4}). 
 
We start with two Lemmas from \citet[Lem. 4.1, Lem. 4.2, (4.12)]{morgan1989ges}  which follow from the results of \citet{ladyzhenskaya1968linear}[Th. 9.1~p.341]. We then use a duality approach to prove global existence of classical solutions to (\ref{eqn:scaled}).
\begin{Lem}[Global existence]\label{Lem:H_imbedding}
Let $\tilde{\myvec{u}}$ be the solution of Problem (\ref{eqn:scaled}). Let the function $H$ fulfil conditions (\ref{G1})---(\ref{G3}) in \S \ref{subsec:lyapunov} and let the polynomial growth restriction on $\tilde{\myvec{f}}$  (\ref{G5_scaled}) hold. Let $T_{\max}$ be as defined in  (\ref{eqn:T_max_defn}) and suppose that,  
\begin{equation}\label{eqn:H_imbedding}
\begin{cases}
\begin{split}
&\text{ for }0\leq\tau<T<T_{\max}\text{ and for all } {p}\in(1,\dotsc,\infty)\\
&\text{ there exist } M_p,N_p>0\text{ and }0<\delta_p<1\text{ such that }\\
&\sum_{i=1}^m\|h_i(\tilde{{u}}_i)\|_{L_p\left(\Omega_0\times(\tau,T)\right)}\leq{M}_p(T-\tau)+N_p(T-\tau)\|H(\tilde{\myvec{u}})\|_{L_p\left(\Omega_0\times(\tau,T)\right)}^{\delta_p},
\end{split}
\end{cases}
\end{equation}
then $T_{\max}=\infty$.
\end{Lem}
\begin{Defn}[Dual problem]\label{defn:dual_pb}
A key ingredient of the proof of Theorem \ref{Thm:existence_time_dependent_f} is the dual solution ${\psi}$. Where for $i=1,\dotsc,m$, ${\psi}$ is the solution of the scalar equation
\begin{equation}\label{eqn:psi_defn}
\begin{split}
\begin{cases}
\partial_t\psi(\myvec{\xi},t)=-D_i\lap\psi(\myvec{\xi},t)-\theta(\myvec{\xi},t) &\text{for }\vec{\xi}\in\Omega_0 \text{ and } 0\leq{t}<T<T_{\max}\\
[\myvec{\nu}\cdot\nabla{\psi}](\myvec{\xi},t)=0,  &\myvec{\xi}\in\partial\Omega_0, t\in[0,T)\\
\psi(\myvec{\xi},T)=0, &\myvec{\xi}\in\Omega_0,
\end{cases}
\end{split}
\end{equation}
where $\theta\geq{0}$ is such that, for all $p\in(1,\dotsc,\infty)$, $\|\theta\|_{L_p(\Omega_0\times[0,T])}=1$.
\end{Defn}
\begin{Lem}[Control of the solution to the dual problem (\ref{eqn:psi_defn})]\label{Lem:psi_regularity}
 Let $\psi$ be as defined in \ref{defn:dual_pb}. For ${i}=1,\dotsc,{m}$  and for $p\in (1,\dotsc,\infty)$, there exists $C_{p,T}>0$ such that,
\begin{align}
\|\psi\|_{L_p\left(\Omega_0;L_{\infty}[0,T]\right)}&\leq{C}_{p,T} \label{eqn:psi_Lpo}
\end{align}
\end{Lem}
We now state the main result of this section. Namely, the applicability of a special case of Theorem \ref{globexis} to solutions of Problem (\ref{eqn:scaled}).
\begin{The}[Sufficient conditions for global existence of solutions to Problem (\ref{eqn:scaled})]\label{Thm:existence_time_dependent_f}
Let Assumptions \ref{rhodefn}, \ref{finv} and \ref{lyap} hold. Let  $H$, $\myvec{f}$ and $r$  satisfy the conditions in \S \ref{subsec:lyapunov} with   ${r}\leq1$ (cf. (\ref{G4})), i.e., Problem (\ref{eqn:fixed})  admits a global classical solution by Theorem \ref{globexis}. Then, Problem (\ref{eqn:scaled}) admits a global classical solution, i.e., $T_{\max}=\infty$ (cf. (\ref{eqn:T_max_defn})).
\end{The}
\begin{Proof}
We proceed by contradiction. Assume $T_{\max}<\infty$. Let $\tilde{\myvec{u}}$ and $\tilde{\myvec{f}}$ (cf. \ref{eqn:f_tilde_defn}) be the solution and zero order term of Problem (\ref{eqn:scaled}) respectively.  From the proof of Lemma \ref{Lem:equivalence_lyap}  the polynomial growth restriction (\ref{G5_scaled}) is satisfied by $\tilde{\myvec{f}}$ and $H$.  Since $T_{\max}<\infty$, Lemma \ref{Lem:H_imbedding} implies that (\ref{eqn:H_imbedding}) does not hold. Let $j\in[1,\dotsc,m]$ denote the smallest $k$ for which  ${\displaystyle\sum_{i=1}^k\|h_i(\tilde{{u}}_i)\|_{L_{p}\left(\Omega_0\times(\tau,T)\right)}}$ does not satisfy (\ref{eqn:H_imbedding}). From the proof of Lemma \ref{Lem:equivalence_lyap}, the intermediate sum condition (\ref{G4_scaled}) is satisfied for with $\tilde{r}=1$ (cf. (\ref{G4_scaled})). From Lemma \ref{Lem:apriori_f_tilde}, we have the a priori estimate  (\ref{eqn:one_localap}). 

We will show  (\ref{eqn:one_localap})  and (\ref{G4_scaled})  imply   (\ref{eqn:H_imbedding}) is satisfied for $j$, obtaining a contradiction. We split the remainder of the proof into steps.

\begin{enumerate}[Step 1:]
\item
We first show the following inequality (corresponding to \citet[(4.6)---(4.9)]{morgan1989ges}):  For $0<T<{T}_{\max}$ 
\begin{equation}\label{eqn:duality_three}
\begin{split}
\int_{\Omega_0}\int_0^T\sum_{i=1}^ja_{ji}\frac{D_i}{D_j}h_i\left(\tilde{u}_i(\myvec{\xi},s)\right)&\theta(\myvec{\xi},s)\dif \myvec{\xi}\dif s\\
\leq&\int_0^T\int_{\Omega_0}\sum_{i=1}^{j-1}a_{ji}\left(1-\frac{D_i}{D_j}\right)h_i\left(\tilde{u}_i(\myvec{\xi},s)\right)\partial_s\psi(\myvec{\xi},s)\\ 
&+\psi(\myvec{\xi},s)\left((k_{1}C_{1}^2 +k_{7}){H}\left(\tilde{\myvec{u}}(\myvec{\xi},s)\right) + k_{8}\right)\dif s\\
&+\sum_{i=1}^ja_{ji}\psi(\myvec{\xi},0)h_i\left(\tilde{u}_i^0(\myvec{\xi})\right)\dif \myvec{\xi}\\
:=&I_1+I_2+I_3.
\end{split}
\end{equation} 
From (\ref{eqn:psi_defn}) we have for $i\in[1,\dotsc,m]$
\begin{equation}
\begin{split}\label{eqn:duality_one}
\int_0^T\int_{\Omega_0}h_i\left(\tilde{u}_i(\myvec{\xi},s)\right)\theta(\myvec{\xi},s)&\dif \myvec{\xi}\dif s\\
=\int_0^T\int_{\Omega_0}&-h_i\left(\tilde{u}_i(\myvec{\xi},s)\right)\left[\partial_s\psi+D_j\lap\psi\right](\myvec{\xi},s)\dif \myvec{\xi}\dif s\\
\leq\int_0^T\int_{\Omega_0}&-h_i\left(\tilde{u}_i(\myvec{\xi},s)\right)\partial_s\psi(\myvec{\xi},s)\\
&-\frac{D_j}{D_i}\psi(\myvec{\xi},s)D_i\lap_{\myvec{\xi}}{h}_i\left(\tilde{u}_i(\myvec{\xi},s)\right)\dif \myvec{\xi}\dif s,\\
\end{split}
\end{equation}
where we have used integration by parts and the homogenous Neumann boundary conditions. From  Problem (\ref{eqn:scaled}) and the convexity of $H$ (\ref{G2}), we have
\begin{equation}
\begin{split}\label{eqn:psi_laph_bound}
-\int_0^T\int_{\Omega_0}\psi(\myvec{\xi},s)D_i\lap_{\myvec{\xi}}{h}_i&\left(\tilde{u}_i(\myvec{\xi},s)\right)\dif \myvec{\xi}\dif s\\
\leq\int_0^T\int_{\Omega_0}&\psi(\myvec{\xi},s)h_i^{\prime}\left(\tilde{u}_i(\myvec{\xi},s)\right)\left(\tilde{f}_i\left(\myvec{\tilde{u}}(\myvec{\xi},s),s\right)-\partial_s\tilde{u}_i(\myvec{\xi},s)\right)\dif \myvec{\xi}\dif s\\
=\int_0^T\int_{\Omega_0}&\psi(\myvec{\xi},s)h_i^{\prime}\left(\tilde{u}_i(\myvec{\xi},s)\right)\tilde{f}_i\left(\myvec{\tilde{u}}(\myvec{\xi},s),s\right)\\
+{h}_i&\left(\tilde{u}_i(\myvec{\xi},s)\right)\partial_s\psi(\myvec{\xi},s)\dif \myvec{\xi}\dif s
+\int_{\Omega_0}\psi(\myvec{\xi},0)h_i\left(\tilde{u}_i^0(\myvec{\xi})\right)\dif \myvec{\xi},
\end{split}
\end{equation}
where we have used integration by parts and the final condition of (\ref{eqn:psi_defn}). Combining (\ref{eqn:psi_laph_bound}) and (\ref{eqn:duality_one}), we obtain 
\begin{equation}
\begin{split}\label{eqn:duality_two}
\int_0^T\int_{\Omega_0}h_i\left(\tilde{u}_i(\myvec{\xi},s)\right)\theta(\myvec{\xi},s)\dif \myvec{\xi}\dif s\leq&\int_{\Omega_0}\int_0^T\left(\frac{D_j}{D_i}-1\right)h_i\left(\tilde{u}_i(\myvec{\xi},s)\right)\partial_s\psi(\myvec{\xi},s)\\
&+\frac{D_j}{D_i}\psi(\myvec{\xi},s)h_i^{\prime}\left(\tilde{u}_i(\myvec{\xi},s)\right)\tilde{f}_i\left(\myvec{\tilde{u}}(\myvec{\xi},s),s\right)\dif s\\
&+ \frac{D_j}{D_i}\psi(\myvec{\xi},0)h_i\left(\tilde{u}_i^0(\myvec{\xi})\right)\dif \myvec{\xi}.
\end{split}
\end{equation}
Summing (\ref{eqn:duality_two}) over $i\leq{j}$ and using condition (\ref{G4_scaled}) with $\tilde{r}=1$, we obtain (\ref{eqn:duality_three}). For the case $j=1$,  we have introduced the convention $\displaystyle{\sum_{i=1}^{0}(\cdot)=0}$.
%%%%%%%%%%%%%%%%%%%%%%%%%%%%%%%%%%%%%%%%%%%%%%%%%%%%%%%%%%%%%%%%%%%%%%%%%%%%%%%%%%%%%%%%%%%%%%%%%%%%%%%%%%%%%%
\item
We shall use Lemma \ref{Lem:psi_regularity} and (\ref{eqn:duality_three}) to obtain the following inequality (as in \citet[(4.10)---(4.16)]{morgan1989ges}): For all $p\in(1,\dotsc,\infty)$
there exists $K_{p,T}>0$ and $0<\delta_p<1$ independent of $\tilde{\myvec{u}}$ and $\theta$ such that, 
\begin{equation}\label{eqn:duality_final}
\int_0^T\int_{\Omega_0}\sum_{i=1}^ja_{ji}\frac{D_i}{D_j}h_i\left(\tilde{u}_i(\myvec{\xi},s)\right)\theta(\myvec{\xi},s)\dif \myvec{\xi}\dif s\leq
K_{p,T}\left(1+\|H\left(\tilde{\myvec{u}}\right)\|^{\delta_p}_{L_p\left(\Omega_0\times[0,T]\right)}\right).
\end{equation}
Let $p,q\in(1,\dotsc,\infty)$ be such that, $\frac{1}{p}+\frac{1}{q}=1$. Dealing firstly with $I_1$ (cf. (\ref{eqn:duality_three})), we have by H\"older's inequality and the regularity estimate (\ref{eqn:psi_Lpo}) 
\begin{equation}\label{eqn:I_1_bound}
\begin{split}
I_1&\leq\sum_{i=1}^{j-1}a_{ji}\left\vert1-\frac{D_i}{D_j}\right\vert{C}_{q,T}T\|h_i\left(\tilde{u}_i\right)\|_{L_p\left(\Omega_0\times[0,T]\right)}\\
&\leq \sum_{i=1}^{j-1}a_{ji}\left\vert1-\frac{D_i}{D_j}\right\vert{C}_{q,T}T\left(M_p+N_p\|H\left(\tilde{\myvec{u}}\right)\|^{\delta_q}_{L_p\left(\Omega_0\times[0,T]\right)}\right),
\end{split}
\end{equation}
where we have the assumption that (\ref{eqn:H_imbedding}) is valid for $i<j$. Dealing with $I_2$, we have by H\"older's inequality and the regularity estimate (\ref{eqn:psi_Lpo})
\begin{equation}\label{eqn:I_2_bound}
\begin{split}
I_2\leq&\int_{\Omega_0}\|\psi(\myvec{\xi},\cdot)\|_{L_{\infty}[0,T]}\left(\int_0^T((k_{1}C_{1}^2 +k_{7}){H}\left(\tilde{\myvec{u}}(\myvec{\xi},s)\right) + k_{8} \dif s\right)\dif \myvec{\xi}\\
\leq&{C}_{q,T}\left((k_{1}C_{1}^2 +k_{7})\left\|\int_0^T{H}\left(\tilde{\myvec{u}}(\cdot,s)\right)\right\|_{L_q(\Omega_0)}+k_8{T}\right)\\
\leq&{C}_{q,T}\left(g(T)\left\vert\Omega_0\right\vert^{1/p}+k_8{T}\right),
\end{split}
\end{equation}
for some $g\in{C}[0,\infty)$. Where we have used   the a priori estimate (\ref{eqn:one_localap}). Finally dealing with $I_3$ using H\"older's inequality and estimate (\ref{eqn:psi_Lpo}) we have
\begin{equation}\label{eqn:I_3_bound}
\begin{split}
I_3\leq&\sum_{i=1}^ja_{ji}{C}_{q,T}\left\|h_i\left(\tilde{u}_i^0\right)\right\|_{L_p\left(\Omega_0\times[0,T]\right)}\\
\leq&\sum_{i=1}^ja_{ji}{C}_{q,T}C_p,
\end{split}
\end{equation}
where we have used the boundedness of $\tilde{\myvec{u}}^0$ and  condition (\ref{G2}). Combining  (\ref{eqn:I_1_bound}), (\ref{eqn:I_2_bound}), (\ref{eqn:I_3_bound}) and (\ref{eqn:duality_three}) yields (\ref{eqn:duality_final}).
%%%%%%%%%%%%%%%%%%%%%%%%%%%%%%%%%%%%%%%%%%%%%%%%%%%%%%%%%%%%%%%%%%%%%%%%%%%%%%%%%%%%%%%%%%%%%%%%%%%%%%%%%%%%%%
\item
We now show (\ref{eqn:H_imbedding}) holds for ${j}$. Let $p,q\in(1,\dotsc,\infty)$, be such that, $\frac{1}{p}+\frac{1}{q}=1$. We recall, from (\ref{G2}) and Definition \ref{defn:dual_pb}, that for $i\in[1,\dotsc,m],$ $h_i,\theta\geq{0}$ and $\|\theta\|_{L_q(\Omega_0\times(0,T))}=1$. Using duality we obtain
\begin{equation}
\begin{split}\label{eqn:duality_imbedding}
\min_{i\leq{j}}\left(a_{ji}\frac{D_i}{D_j}\right)\sum_{i=1}^j\|h_i(\tilde{{u}}_i)&\|_{L_p\left(\Omega_0\times(0,T)\right)}\\
=&\min_{i\leq{j}}\left(a_{ji}\frac{D_i}{D_j}\right)\int_0^T\int_{\Omega_0}\sum_{i=1}^jh_i\left(\tilde{u}_i(\myvec{\xi},s)\right)\theta(\myvec{\xi},s)\dif \myvec{\xi}\dif s\\
\leq&\int_0^T\int_{\Omega_0}\sum_{i=1}^ja_{ji}\frac{D_i}{D_j}h_i\left(\tilde{u}_i(\myvec{\xi},s)\right)\theta(\myvec{\xi},s)\dif \myvec{\xi}\dif s\\
\leq& K_{p,T}\left(1+\|H\left(\tilde{\myvec{u}}\right)\|^{\delta_p}_{L_p\left(\Omega_0\times[0,T]\right)}\right),
\end{split}
\end{equation}
where we have used  (\ref{eqn:duality_final}). 
\end{enumerate}
Thus  we have a contradiction and  we conclude $T_{\max}=\infty$ completing the proof.
\end{Proof}

The proof of Theorem \ref{globexis} follows from  more technical use of H\"older's inequality in (\ref{eqn:I_2_bound}) and in the case of assertion (\ref{eqn:globexis_two})  we also require the a priori estimate (\ref{eqn:two_localap}). We refer to \citet[(4.13)---(4.16) and (4.27)---(4.19)]{morgan1989ges} for specific details.

}

\end{document}